\documentclass{article} 

\usepackage[final]{graphicx}
\usepackage{amsmath,amsfonts,amssymb}
\usepackage{verbatim}
\usepackage{color}
\usepackage{epsfig,wrapfig}
\usepackage[numbers,sort&compress,comma]{natbib}
\usepackage[draft=false,letterpaper,breaklinks,colorlinks,linktocpage,citecolor=blue,linkcolor=blue,urlcolor=blue]{hyperref}
\usepackage[letterpaper, top=1.5in, bottom=1.5in, inner=1.4in, outer=1.4in, foot=.25in]{geometry} 

\usepackage{algorithm}
\usepackage[noend]{algorithmic}

\newcommand{\A}{\mathcal{A}}
\newcommand{\X}{\mathcal{X}}
\newcommand{\GP}{\mbox{GP}}
\newcommand{\y}{\mathbf{y}}
\newcommand{\x}{\mathbf{x}}
\newcommand{\f}{\underline{f}}

\newcommand{\corr}{\mbox{corr}}

\newcommand{\figdir}{figs}

\newcommand{\presec}{\vspace*{-0pt}} 
\newcommand{\postsec}{\vspace*{-0pt}} 
\newcommand{\pressec}{\vspace*{-0pt}} 
\newcommand{\postssec}{\vspace*{-0pt}} 
\newcommand{\prepar}{\vspace*{-0pt}}
\newcommand{\precap}{\vspace*{-0in}}
\newcommand{\postcap}{\vspace*{-0.2in}}
\newcommand{\preeq}{\vspace*{-0pt}}
\newcommand{\posteq}{\vspace*{-0pt}}
\newcommand{\algspace}{\vspace*{0pt}}

\title{Multiresolution Gaussian Processes}
\author{
Emily B.~Fox \\
Department of Statistics\\
University of Washington\\
Seattle, WA 98195 \\
\texttt{ebfox@uw.edu} \\
\and
David B.~Dunson \\
Department of Statistical Science \\
Duke University\\
Durham, NC 27708 \\
\texttt{dunson@stat.duke.edu} \\
}

\begin{document}
\maketitle
\begin{abstract}
We propose a multiresolution Gaussian process to capture long-range, non-Markovian dependencies while allowing for abrupt changes.  The multiresolution GP hierarchically couples a collection of smooth GPs, each defined over an element of a random nested partition.  Long-range dependencies are captured by the top-level GP while the partition points define the abrupt changes.  Due to the inherent conjugacy of the GPs, one can analytically marginalize the GPs and compute the conditional likelihood of the observations given the partition tree.  This property allows for efficient inference of the partition itself, for which we employ graph-theoretic techniques.  We apply the multiresolution GP to the analysis of Magnetoencephalography (MEG) recordings of brain activity.
\end{abstract}
%
%
\presec
\section{Introduction}
\label{sec:intro}
\postsec
\vspace{-0.0in}
A key challenge in many time series applications is capturing long-range dependencies for which Markov-based models are insufficient.  One method of addressing this challenge is through employing a Gaussian process (GP) with an appropriate (non-band-limited) covariance function.  However, GPs typically assume smoothness properties of the underlying function being modeled that can blur key elements of the signal if abrupt changes occur. The Mat\'{e}rn kernel enables less smooth functions, but assumes a stationary process that does not adapt to varying levels of smoothness. Likewise, a changepoint~\cite{Saatci:10} or partition~\cite{Gramacy:08} model between smooth functions fails to capture long range dependencies spanning changepoints. 

Another long-memory process is the fractional ARIMA process~\cite{Diebold:89}, with extensions to infinite variance innovations in~\cite{Kokoszka:96}; however, the appropriateness and robustness of such models for real data analysis has been questioned~\cite{Gorst-Rasmussen:12}.  Wavelet methods have also been proposed, including recently for smooth functions with discontinuities~\cite{Beran:12}.  Such methods inherit the properties and limitations of wavelet analysis, for example, lack of shift invariance.  We take a fundamentally different approach based on GPs that allows (i) direct interpretability, (ii) local stationarity, (iii) irregular grids of observations, and (iv) sharing information across related time series. 

\begin{figure}[t!]
	\centering 
	\hspace{-0.1in}
	\begin{tabular}{cccc} 
		\hspace{-0.5in}
		\includegraphics[height = 0.95in]{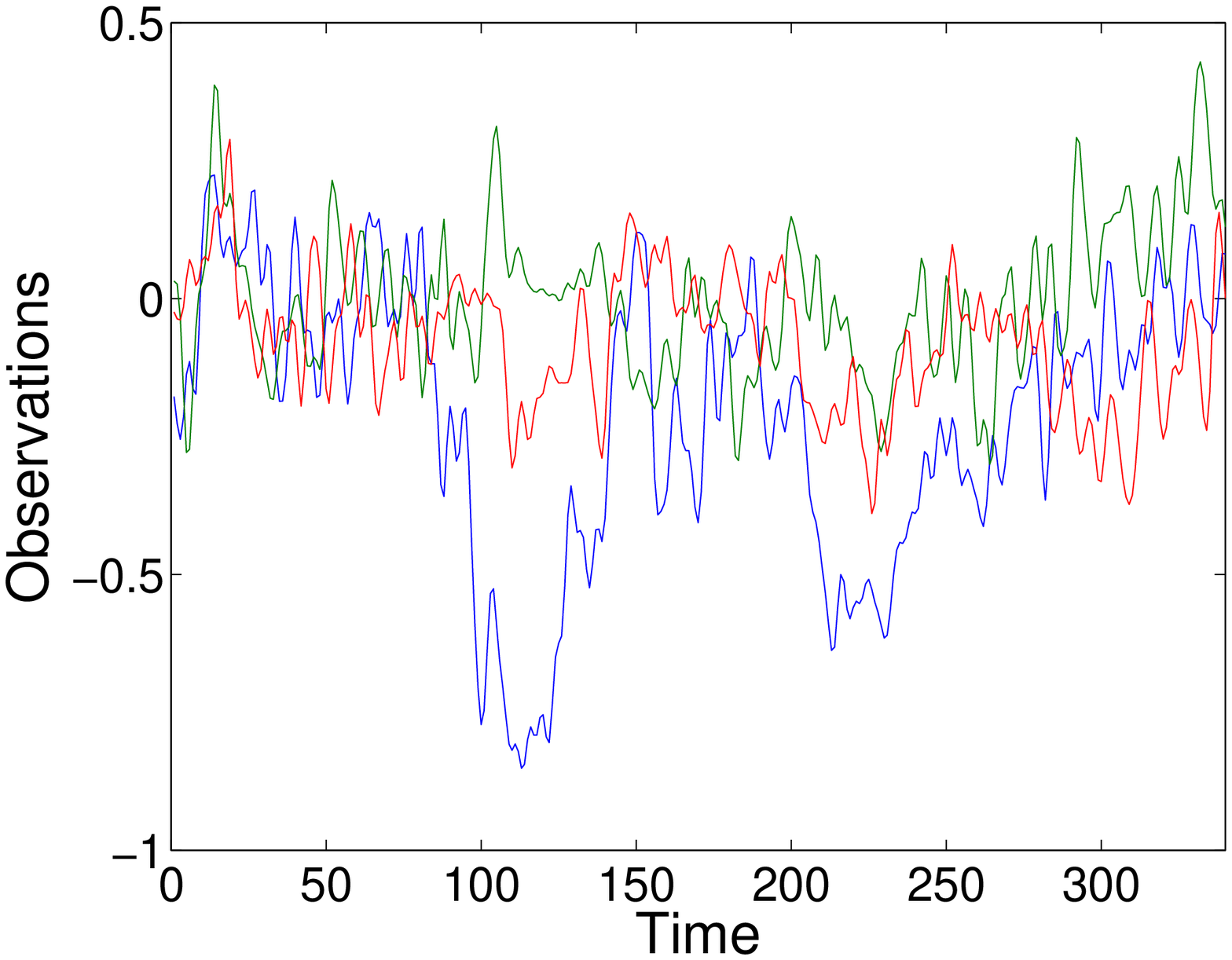} & \hspace{-0.1in} 
		\includegraphics[height = 0.9in]{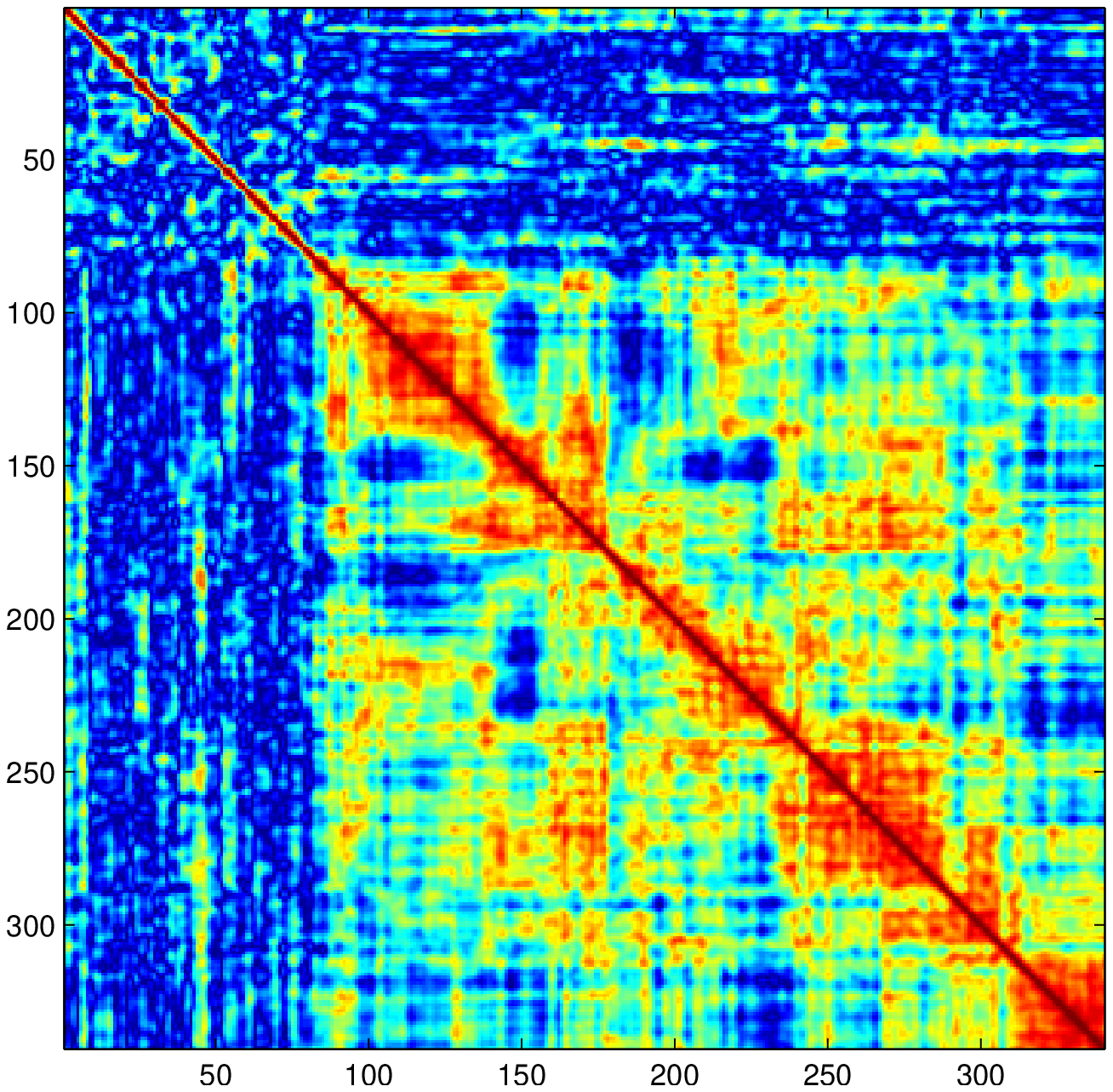} & \hspace{-0.1in}
		\includegraphics[height = 0.9in]{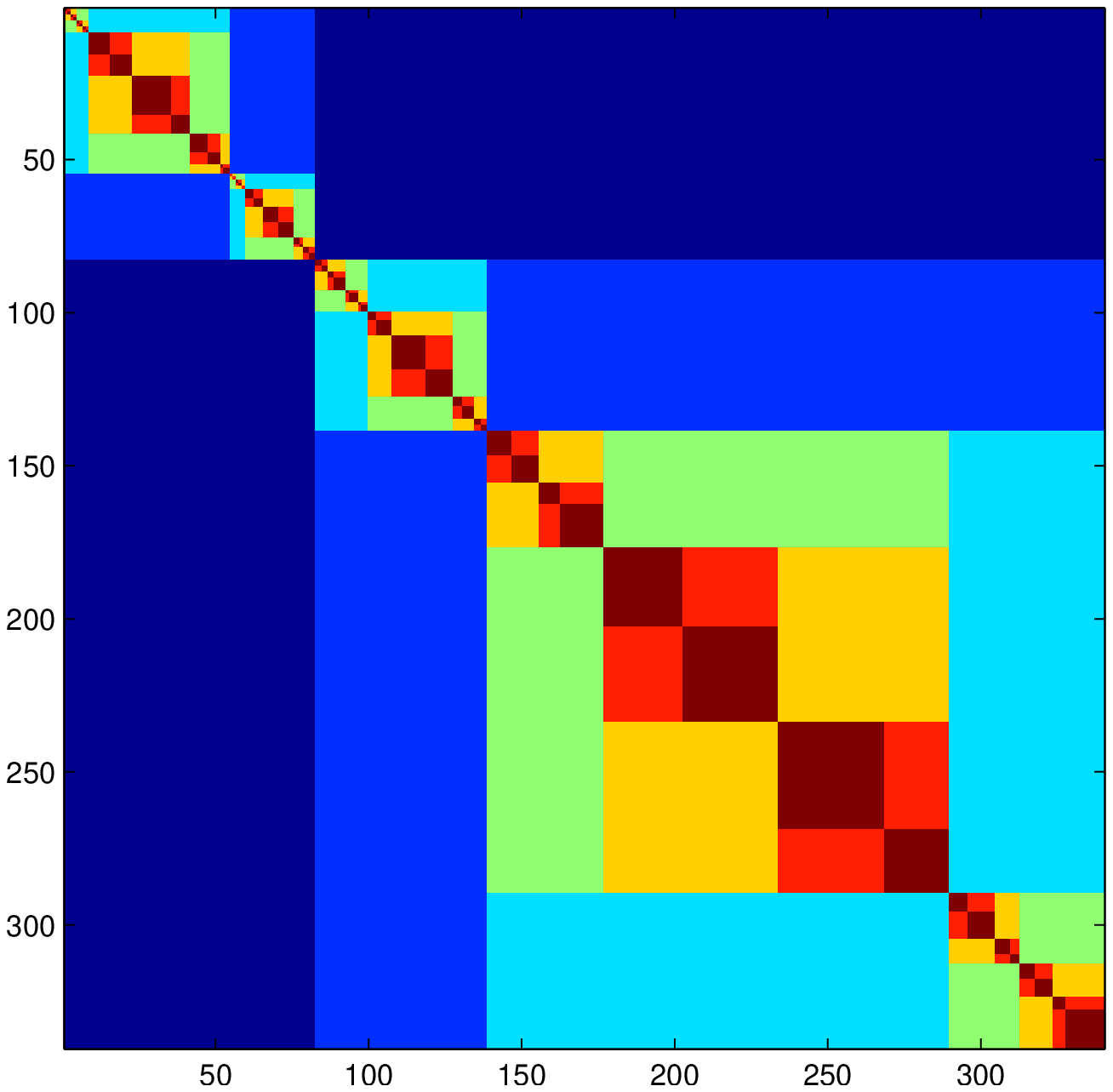} & \hspace{0.4in}
		\includegraphics[width=1.1in]{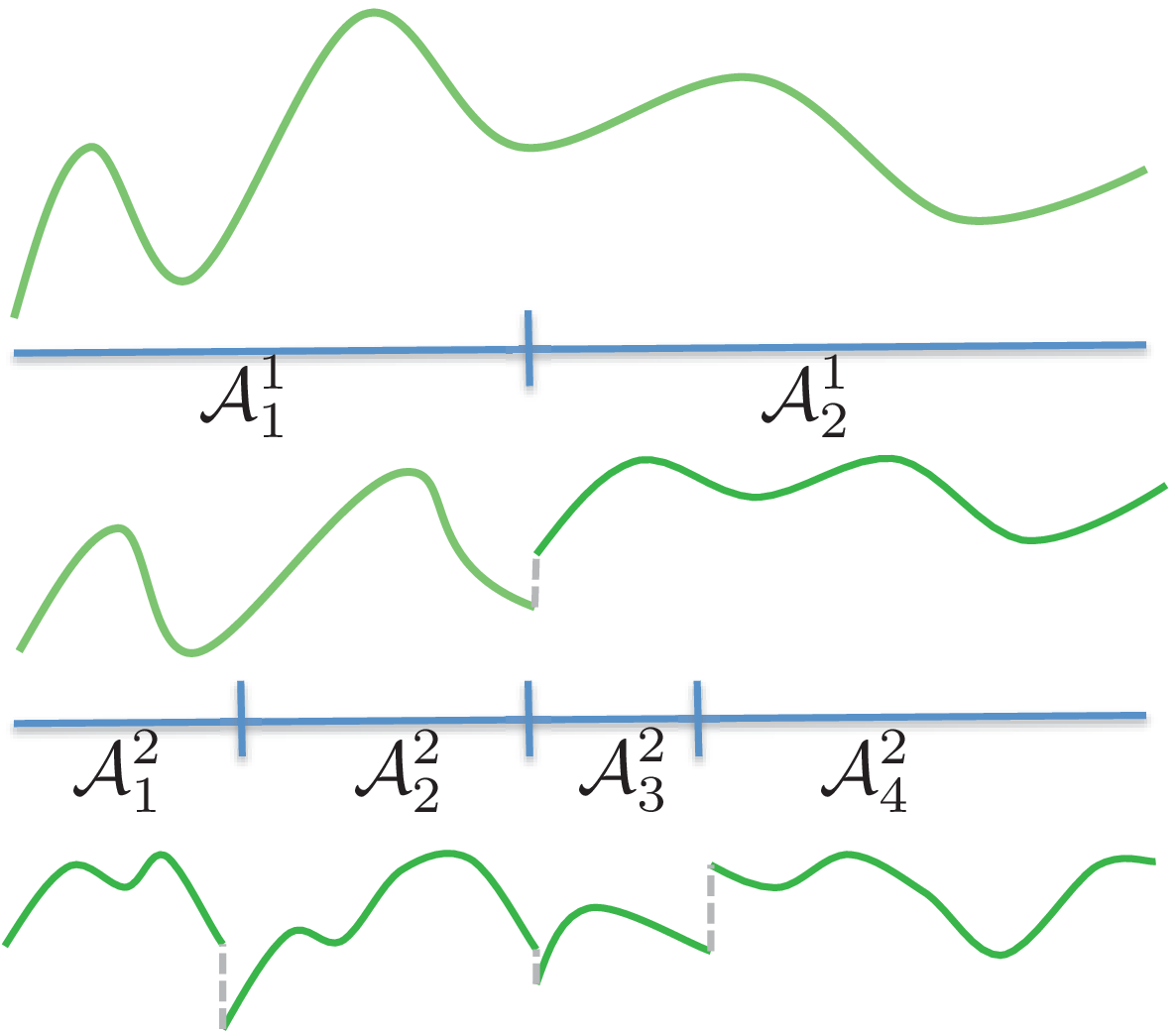}
	\end{tabular} \vspace{-0.1in}
	\hspace{-0.2in}\begin{minipage}[h]{3.25in} 
		\precap\vspace{2pt}
			\caption{\small For sensor 1 and word \emph{house}, \emph{Left:} Data from three trials; \emph{Middle:} Correlation matrix estimated from 20 trials; \emph{Right:}  Hierarchical segmentation produced by recursive minimization of normalized cut objective, with color indicating tree level.} \label{fig:MEGexample} \postcap \vspace{0.1in}
	\end{minipage} 
	\hfill
	\begin{minipage}[h]{2.15in} 
		\vspace*{-6pt}
		\caption{\small mGP: Parent function is split by $\A^1 = \{\A^1_1,\A^1_2\}$.  Recursing down the tree, each partition has a GP with mean given by its parent function restricted to that set.} \label{fig:multiResGP}
		\postcap \vspace{0.1in}
	\end{minipage}
\end{figure}

As a motivating application, consider Magnetoencephalography (MEG) recordings of brain activity in response to some word stimulus.  Due to the low signal-to-noise-ratio (SNR) regime, multiple trials are often recorded, presenting a \emph{functional data analysis} scenario.  Each trial results in a noisy trajectory with key discontinuities (e.g., after stimulus onset).  Although there are overall similarities between the replicates, there are also key differences that occur based on various physiological phenomena, as depicted in Fig.~\ref{fig:MEGexample}.  We clearly see abrupt changes as well as long-range correlations.  Key to the data analysis is the ability to share information about the overall trajectory between the single trials without forcing unrealistic smoothness assumptions on the single trials themselves.

In order to capture both long-range dependencies and potential discontinuities, we propose a multiresolution GP (mGP) that hierarchically couples a collection of smooth GPs, each defined over an element of a nested partition set.  Due to the inherent conjugacy of the GPs, one can analytically marginalize the GPs and compute the marginal likelihood of the observations \emph{conditioned on the partition tree}.  This conditional likelihood is equivalent to the likelihood under a GP with a partition-dependent covariance function.  The covariance function encodes local smoothness of the function and enables 
discontinuities at the partition points.  The amount of correlation between any two observations $y_i$ and $y_j$ generated by the mGP at locations $x_i$ and $x_j$ is a function of the distance $||x_i-x_j||$ and which tree levels contain both $x_i$ and $x_j$.  The tree thus encodes non-stationarity. 

The fact that there is an analytic form for the conditional distribution of the observations given the partition allows for efficient importance sampling of the partition itself.  For our proposal distribution, we borrow the idea of \emph{normalized cuts}~\cite{Shi:00} from the theoretical computer science and computer vision communities.  We also present an MCMC sampler incorporating both local and global moves. 
Our inferences integrate over the partition tree, allowing blurring of discontinuities and producing functions which can appear smooth when discontinuities are not present in the data.

%
\presec
\section{Background}
\label{sec:background}
\postsec
%
A GP provides a distribution on real-valued functions $f: \X \rightarrow \Re$, with the property that the function evaluated at any finite collection of points is jointly Gaussian.  The GP, denoted $\mbox{GP}(m,c)$, is uniquely defined by its \emph{mean function} $m$ and \emph{covariance function} $c$.  That is, $f \sim \mbox{GP}(m,c)$ if and only if for all $n\geq 1$ and $x_1,\dots,x_n$, $(f(x_1),\dots,f(x_n)) \sim N_n(\mu,K)$,
with $\mu = [m(x_1), \dots, m(x_n)]$ and $[K]_{ij} = c(x_i,x_j)$.  The properties (e.g., continuity, smoothness, periodicity, etc.) of functions drawn from a given GP are determined by the covariance function.  The squared exponential kernel, $c(x,x') = d\exp(-\kappa ||x-x'||_2^2)$, leads to smooth functions. Here,
$d$ is a \emph{scale} hyperparameter and $\kappa$ is the \emph{bandwidth} determining the extent of the correlation in $f$ over $\mathcal{X}$.  See~\cite{RasmussenWilliams} for further details.
\presec
\section{Multiresolution Gaussian Process Formulation}
\label{sec:model}
\postsec
%
Our interest is in modeling a function $g$ that (i) is locally smooth, (ii) exhibits long-range correlations (i.e., $\corr(g(x),g(x'))>0$ for $||x-x'||$ relatively large), and (iii) has abrupt changes.  We begin by modeling a single function, but with a specification that readily lends itself to modeling a \emph{collection} of functions that share a similar global trajectory, as explored in Sec.~\ref{sec:multiple}.  
\prepar
\paragraph{Generative Model}
Assume a set of noisy observations $y = \{y_1,\dots,y_n\}$, $y_i \in \Re$, of the function $g$ at locations $\{x_1,\dots,x_n\}$, $x_i \in \X\subset \Re^p$:
\preeq
\begin{align}
	y_i = g(x_i) + \epsilon_i, \quad \epsilon_i \sim N(0,\sigma^2).
	\label{eqn:obsModel}
	\posteq
\end{align}
To capture both long-range dependencies and abrupt changes, we hierarchically define the function $g$ as follows.  Let $\A=\{\A^0,\A^1,\dots,\A^{L-1}\}$ be a set of nested partitions of $\mathcal{X}$ with $\A^0 = \mathcal{X}$ and binary splits such that $\mathcal{X} = \bigcup_i \A^\ell_i$ and $\A^{\ell-1}_i = \A^{\ell}_{2i-1} \cup \A^{\ell}_{2i}$. We define a \emph{global parent function} on $\A^0$ as $f^0 \sim \GP(0,c^0)$. Then, over each partition set $\A^{\ell}_i$ of the binary tree we define
\begin{align}
	f^{\ell}(\A^{\ell}_i) \sim \GP(f^{\ell-1}(\A^{\ell}_i),c^{\ell}_i).
\end{align}
That is, the mean of the GP is given by the parent function restricted to the current partition set.  Finally, we set $g = f^{{L-1}}$.  A pictorial representation of the mGP is shown in Fig.~\ref{fig:multiResGP}.  
%
%
\prepar
\paragraph{Covariance Function}
%
We assume a squared exponential kernel $c^{\ell}_i = d^{\ell}_i\exp(-\kappa^{\ell}_i ||x-x'||_2^2)$, encouraging local smoothness over each partition set $\A_i^\ell$.  Allowing $d^{\ell}_i=0$ enables unbalanced trees---the child function exactly equals the parent function.  However, we focus on $d^{\ell}_i=d^{\ell}$ with $\sum_{\ell=1}^\infty (d^{\ell})^2 <1$ for finite variance regardless of tree depth and additionally encouraging lower levels to vary less from their parent function, providing regularization and robustness to the choice of $L$.

We typically assume bandwidths $\kappa^{\ell}_i = \kappa/||\A^{\ell}_i||_2^2$ so that each child function is locally as smooth as its parent.  One can think of this formulation as akin to a fractal process: zooming in on any partition, the locally defined function has the same smoothness as that of its parent over the larger partition. Thus, lower levels encode finer-resolution details. We denote the covariance hyperparameters as $\theta = \{d^0,\dots,d^{L-1},\kappa\}$, and omit the dependency in conditional distributions for notational simplicity.
\prepar
\paragraph{Induced Marginal GP}
The conditional independencies of our mGP imply that
\preeq 
\begin{align}
	p(g \mid \A) = \int p(f^0)\prod_{\ell=1}^{L-1} p(f^{\ell}\mid f^{\ell-1},\A^\ell) df^{0:L-2}.
	\label{eqn:marg_int}
	\posteq
\end{align}
Due to the inherent conjugacy of the GPs, one can analytically marginalize the hierarchy of GPs \emph{conditioned on the partition tree} $\A$ yielding
\preeq 
\begin{align}
	g \mid \A \sim \GP(0,c^*_{\A}), \quad c^*_{\A} = \sum_{\ell=0}^{L-1} \sum_i c^\ell_i \mathbb{I}_{\A^\ell_i}.
	\label{eqn:inducedGP}
	\posteq
\end{align}
Here, $\mathbb{I}_{\A^\ell_i}(x,x') = 1$ if $x,x' \in \A^{\ell}_i$ and 0 otherwise.  Eq.~\eqref{eqn:inducedGP} provides an interpretation of the mGP as a (marginally) partition-dependent GP, where the partition $\A$ defines the discontinuities in the covariance function $c^*_{\A}$.  The covariance function encodes local smoothness of $g$ and discontinuities at the partition points.  Note that $c^*_{\A}$ defines a \emph{non-stationary} covariance function.

The correlation between any two observations $y_i$ and $y_j$ at locations $x_i$ and $x_j$ generated as in Eq.~\eqref{eqn:obsModel} is a function of how many tree levels contain both $x_i$ and $x_j$ and the distance $||x_i-x_j||$.  Let $r_i^\ell$ index the partition set such that $x_i \in \A_{r_i^\ell}^\ell$ and $L_{ij}$ be the lowest level for which $x_i$ and $x_j$ fall into the same set (i.e., the largest $\ell$ such that $r_i^\ell = r_j^\ell$).  Then, for $x_i \neq x_j$
%
\preeq
\begin{align}
	\corr(y_i,y_j \mid \A) &=
	\frac{\sum_{\ell=0}^{L_{ij}} c_{r_i^{\ell}}^\ell(x_i,x_j)}{\sqrt{(\sigma^2 + \sum_{\ell=0}^{L-1} c_{r_i^{\ell}}^\ell(x_i,x_i))(\sigma^2 + \sum_{\ell=0}^{L-1} c_{r_j^{\ell}}^\ell(x_j,x_j))}}\\
	&= \frac{\sum_{\ell=0}^{L_{ij}} d^\ell\exp(-\kappa||x_i-x_j||_2^2/||\A_{r_i^\ell}^\ell||_2^2)}{\sigma^2 + \sum_{\ell=0}^{L-1} d^\ell},
	\posteq
\end{align}
where the second equality follows from assuming the previously described kernels.
An example correlation matrix is shown in Fig.~\ref{fig:sim}(c).  $\kappa$ determines the width of the bands while $d^{\ell}$ controls the contribution of level $\ell$.  Since $d^{\ell}$ is square summable, lower levels are less influential.  
\prepar
\paragraph{Marginal Conditional Likelihood}
Based on a \emph{vector} of observations $\y=[y_1 \cdots y_n]'$ at locations $\x=[x_1 \cdots x_n]'$, we can restrict our attention to evaluating the GPs at $\x$.  Let $f^{\ell}(\x) = [f^{\ell}(x_1) \cdots f^{\ell}(x_n)]'$. By definition of the GP, we have
\begin{align}
	 f^{\ell}(\x) \mid f^{\ell-1}(\x),\A^\ell \sim N(f^{\ell-1}(\x),K_\ell), \quad 
	[K_\ell]_{i,j} = \left\{\begin{array}{cc} c^{\ell}_r(x_i,x_j) & x_i,x_j \in \A^{\ell}_r\\
									      0 & otherwise\end{array}\right.,
\end{align}
where the covariance matrix $K_\ell$ is a function of the level-specific partition $\A^\ell$. 
Likewise, observations are generated as $\y \mid g(\x) \sim N(g(\x),\sigma^2I_n)$.  Recalling Eq.~\eqref{eqn:marg_int}, standard results yield
\begin{align}
	g(\x) \mid \A \sim N\bigg(0,\sum_{\ell=0}^{L-1} K_\ell\bigg) \quad \y \mid \A \sim N\bigg(0,\sigma^2I_n + \sum_{\ell=0}^{L-1} K_\ell\bigg).
	\label{eqn:margLike_singleObs}
\end{align}
This result can also be derived from the induced mGP of Eq.~\eqref{eqn:inducedGP}.  Again note that the marginal conditional likelihood is equivalent to the marginal likelihood under a GP with a partition-dependent covariance matrix.  Alternatively, one can condition on the GP at any level $\ell'$:
\begin{align}
	\y \mid f^{\ell'}(\x),\A \sim N\bigg(f^{\ell'}(\x),\sigma^2I_n + \sum_{\ell=\ell'+1}^{L-1} K_\ell\bigg).
	\label{eqn:condLike_singleObs}
\end{align}
A key advantage of the mGP is the conditional conjugacy of the latent GPs that allows us to compute the likelihood of the data simply conditioned on the nested partition $\A$.  This fact is fundamental to the efficiency of the partition inference procedure described in Sec.~\ref{sec:partition}.
\presec
\section{Multiple Trials}
\label{sec:multiple}
\postsec
In many applications, such as the motivating MEG application, one has a \emph{collection} of observations of an underlying signal.  
We refer to the replicates as \emph{trials}.  In order to capture the common global trajectory of these trials while still allowing for trial-specific variability, we model each trial as a realization from an mGP with a \emph{shared} parent function $f^{0}$.  One could trivially allow for alternative structures of hierarchical sharing beyond $f^{0}$ if an application warranted.  For simplicity, and due to the motivating MEG application, we additionally assume shared changepoints between the trials, though this assumption can also be relaxed. 
\prepar
\paragraph{Generative Model} 
For each trial $\y^{(j)} = \{y^{(j)}_1,\dots,y^{(j)}_n\}$, we model
\preeq
\begin{align}
\hspace{-1.5in}y^{(j)}_i = g^{(j)}(x_i) + \epsilon^{(j)}_i, \quad \epsilon^{(j)}_i \sim N(0,\sigma^2),
	\posteq
\end{align}
\begin{wrapfigure}{r}{.26\textwidth}
	\vspace{-0.7in}
\centering
\includegraphics[width=.2\textwidth]{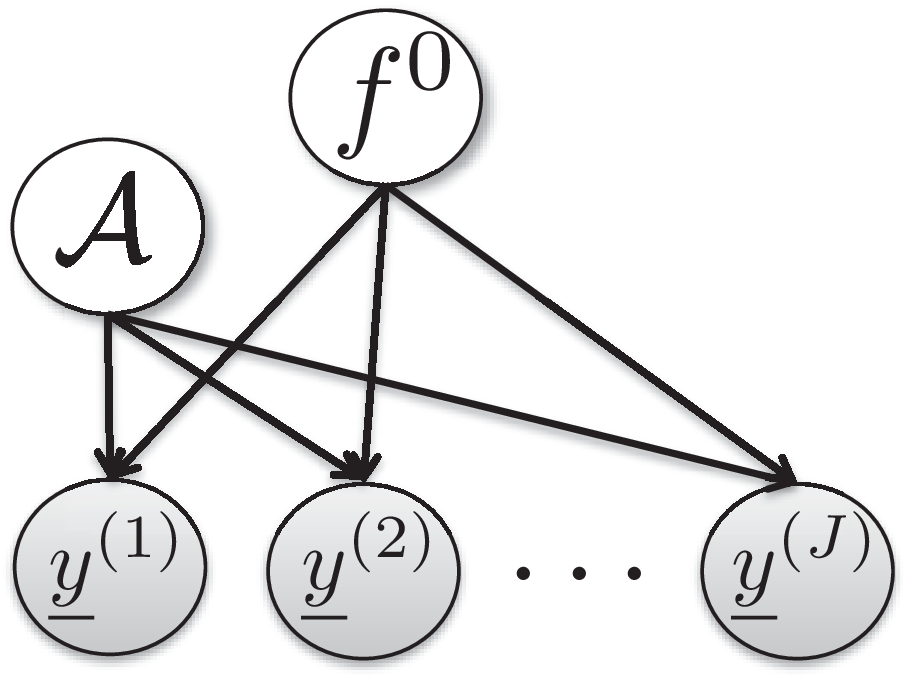}
\vspace*{-6pt}
\caption{\small Graphical model of $J$ replicates from a shared global trajectory $f^0$ and hierarchical partition $\A$.} \label{fig:multTrials}
\vspace*{-6pt}
\end{wrapfigure}
with $g^{(j)} = f^{{L-1},(j)}$ generated from a trial-specific GP hierarchy $\mbox{$f^{0} \rightarrow f^{1,(j)} \rightarrow \cdots \rightarrow f^{{L-1},(j)}$}$ with shared parent $f^{0}$. (Again, alternative structures can be considered.)  Equivalently, from Eq.~\eqref{eqn:condLike_singleObs} with $\ell'=0$, and independently for each $j$
\preeq\preeq
\begin{align}
	\y^{(j)} \mid f^{0}(\x),\A \sim N\bigg(\y^{(j)};f^{0}(\x),\sigma^2I_n+\sum_{\ell=1}^{L-1} K_\ell\bigg),
	\posteq\posteq
\end{align}
See Fig.~\ref{fig:multTrials} for a graphical model.  Note that with our GP-based formulation, we need not assume coincident observation locations $x_1,\dots,x_n$ between the replicates.  However, for simplicity of exposition, we consider shared locations.  We compactly denote the covariance by $\Sigma = \sigma^2I_n+\sum_{\ell=1}^{L-1} K_\ell$.

\begin{figure}[t!]
	\centering
	\begin{tabular}{ccccc} 
		\hspace{-0.2in}
		\includegraphics[width = 1.2in]{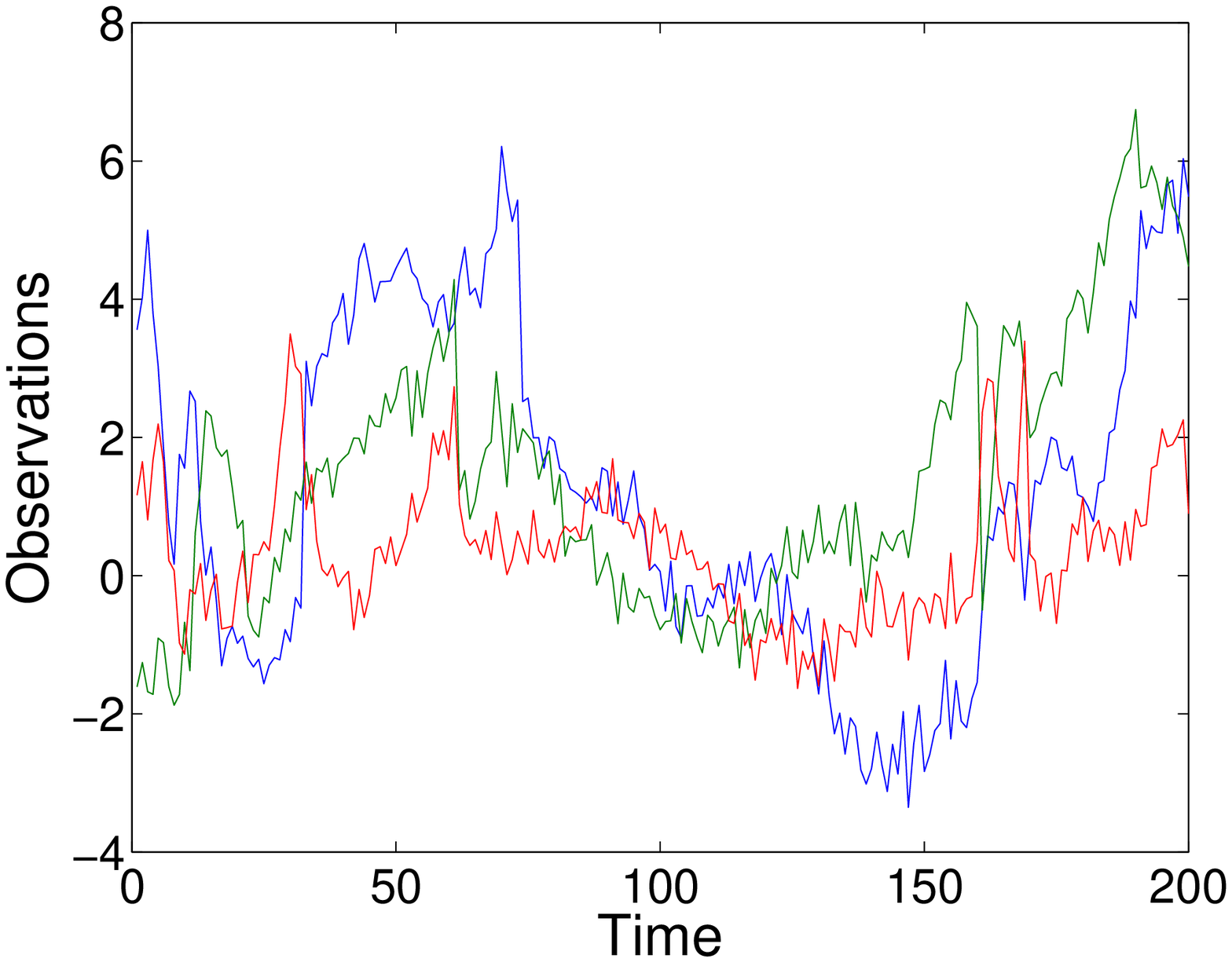} & \hspace{-0.1in} 
		\includegraphics[height = 1in]{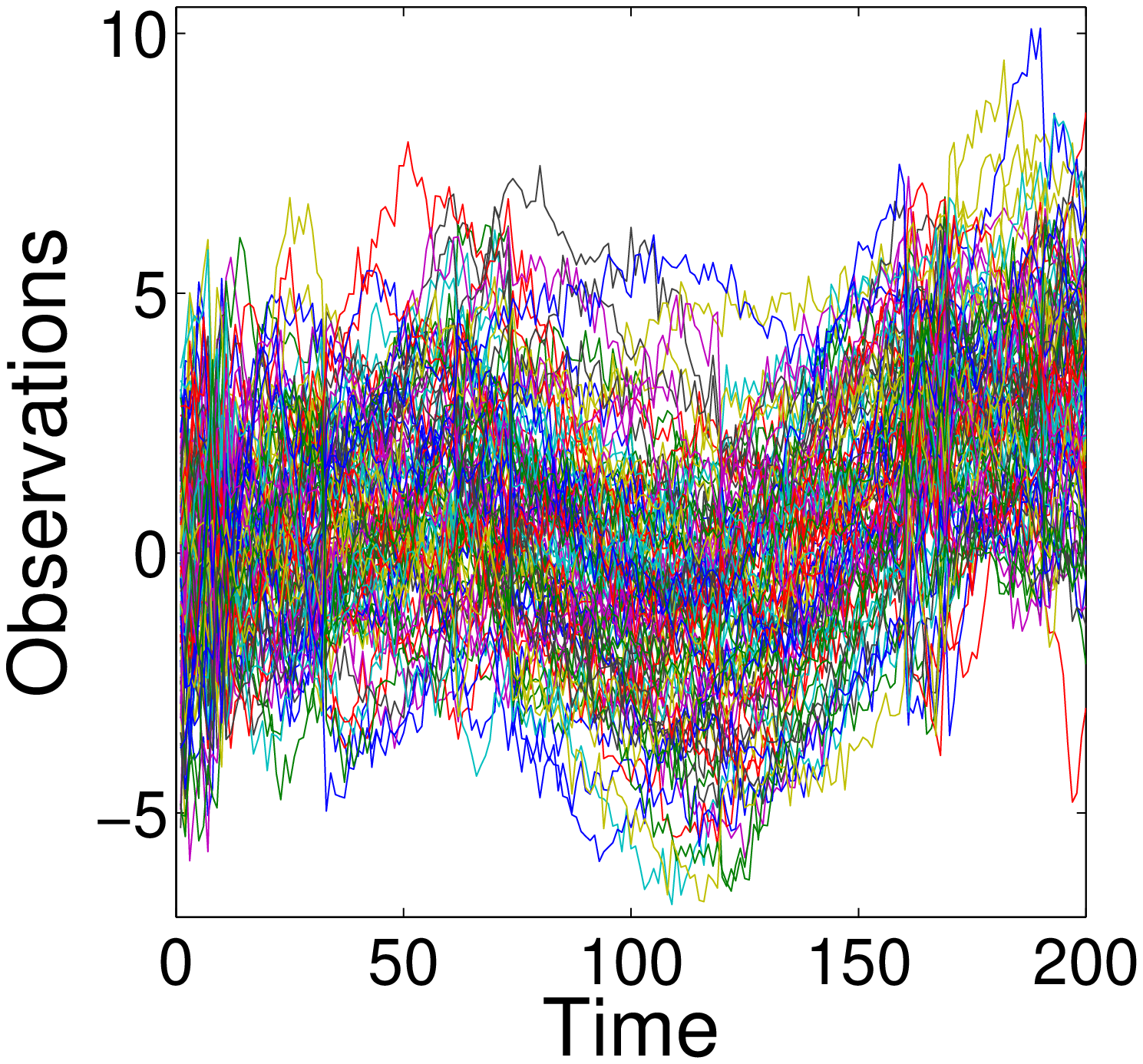} & \hspace{-0.1in} 
		\includegraphics[width = 1in]{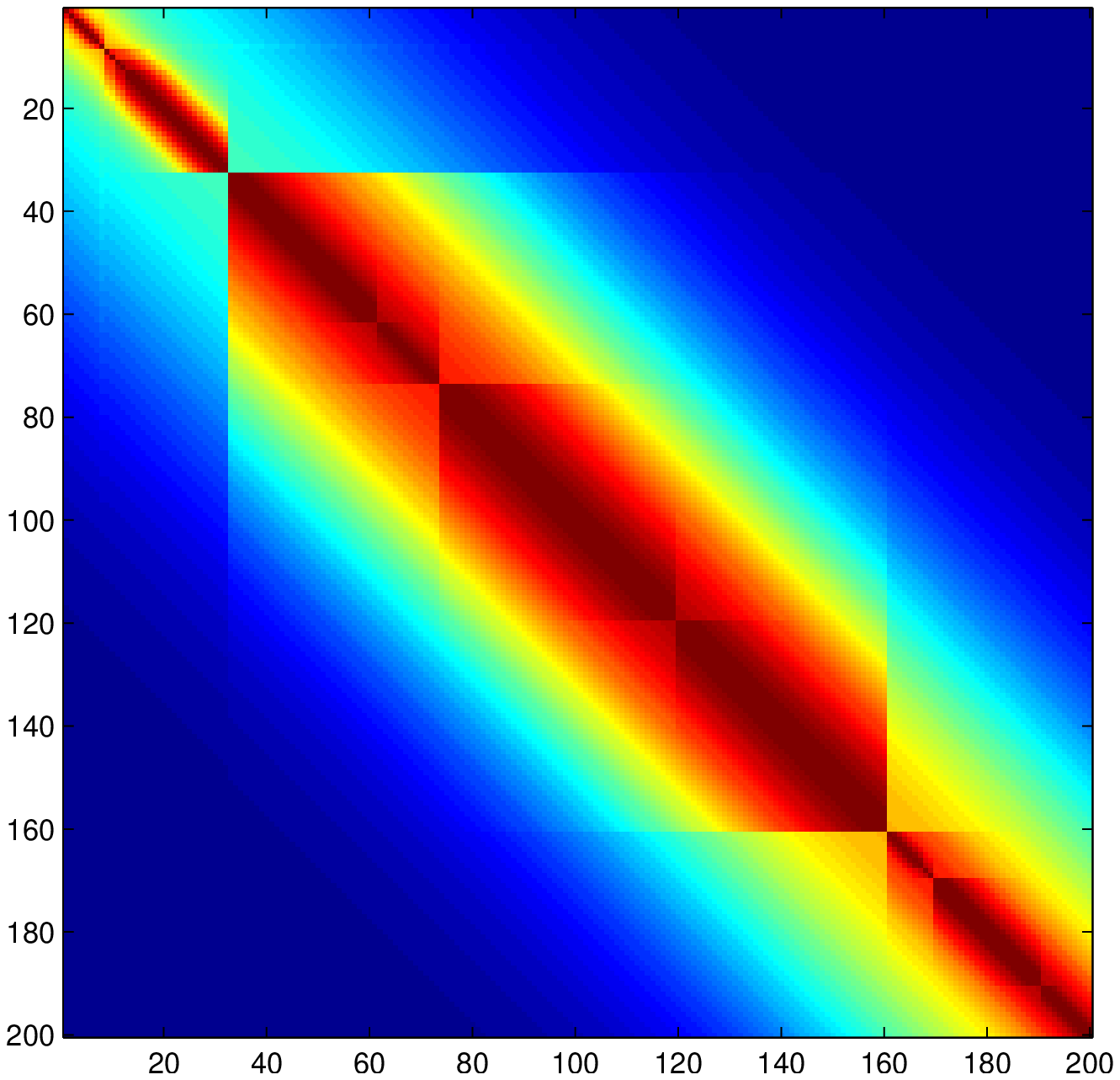} & \hspace{-0.1in}
		\includegraphics[width = 1in]{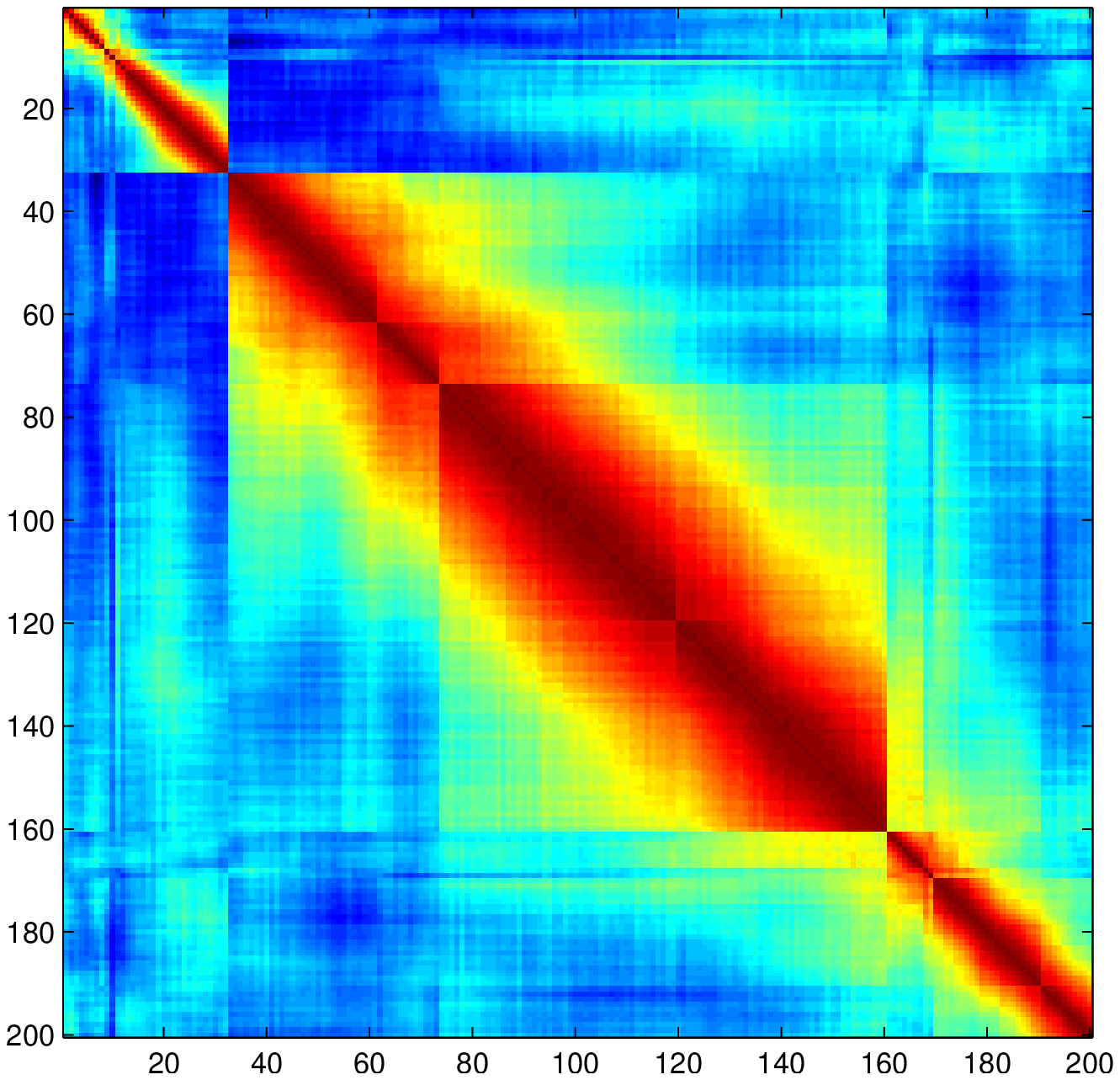} & \hspace{-0.1in}
		\includegraphics[width = 1in]{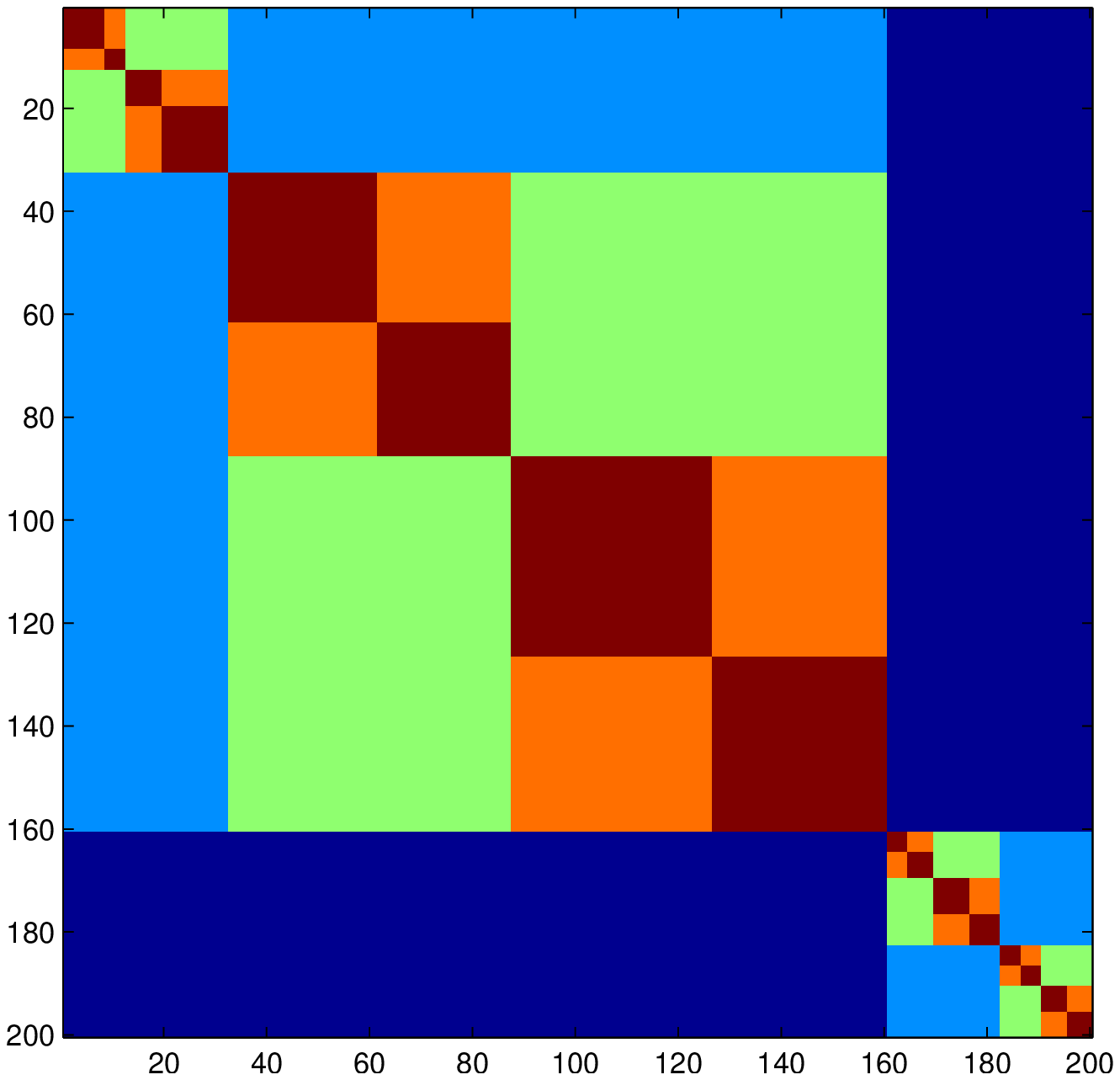}\vspace{-0.05in}\\
		\hspace{-0.2in} {\small (a)} & \hspace{-0.1in} {\small (b)} & \hspace{-0.1in}{\small (c)} & \hspace{-0.1in}{\small (d)} & \hspace{-0.1in} {\small (e)} \vspace{-0.075in}
	\end{tabular}
	\precap
	\caption{\small (a) Three trials and (b) all 100 trials of data generated from a 5-level mGP with a shared parent function $f^0$ and partition $\A$ (randomly sampled). (c) True correlation matrix. (d) Correlation matrix estimated from 100 trials. (e) Hierarchical segmentation produced by recursive minimization of normalized cut objective.
	} \label{fig:sim} \postcap \vspace{0.05in}
\end{figure}

Simulated data generated from a 5-level mGP with shared $f^0$ and $\A$ are shown in Fig.~\ref{fig:sim}.  The sample correlation matrix is also shown.  Compare with the MEG data of Fig.~\ref{fig:MEGexample}.  Both the qualitative structure of the raw time series as well as blockiness of the correlation matrix have striking similarities.
\vspace{-0.1in}
\prepar
\paragraph{Posterior Global Trajectory and Predictions}
Based on a set of trials $\{\y^{(1)},\dots,\y^{(J)}\}$, it is of interest to infer the posterior of $f^{0}$.  Standard Gaussian conjugacy results imply that
\preeq
\begin{align}
	p(f^{0}(\x) \mid \y^{(1)},\dots,\y^{(J)},\A) = N\bigg(\left(K_0^{-1} + J\Sigma^{-1}\right)^{-1}\tilde{\y}, \left(K_0^{-1} + J\Sigma^{-1}\right)^{-1}\bigg),
	\posteq\posteq
\end{align}
where $\tilde{\y} = \Sigma^{-1}\sum_i \y^{(i)}$. Likewise, the predictive distribution of data from a new trial is
\preeq
\begin{align}
	p(\y^{(J+1)} \mid \y^{(1)},\dots,\y^{(J)},\A) &=\int p(\y^{(J+1)} \mid f^{0}(\x),\A)p(f^{0}(\x) \mid \y^{(1)},\dots,\y^{(J)},\A)df^{0}\nonumber\\
	&\hspace{-0.5in}= N\bigg(\left(K_0^{-1} + J\Sigma^{-1}\right)^{-1}\tilde{\y}, \Sigma + \left(K_0^{-1} + J\Sigma^{-1}\right)^{-1}\bigg).
	\label{eqn:preddist}
	\posteq\posteq
\end{align}
\vspace{-0.15in}
\prepar
\paragraph{Marginal Conditional Likelihood}
Since the set of trials $Y = \{\y^{(1)},\dots,\y^{(J)}\}$ are generated from a shared parent function $f^0$, the marginal likelihood does not decompose over trials.  Instead,
\preeq
\begin{align}
	p(Y\mid \A) = \frac{|K_0|^{-1/2}|\Sigma|^{-J/2}}{ (2\pi)^{-nJ/2}|J\Sigma^{-1} + K_0^{-1}|^{1/2}} \exp\bigg(-\frac{1}{2} \sum_i \y^{(i)'}\Sigma^{-1}\y^{(i)} + \frac{1}{2}\tilde{\y}' (J\Sigma^{-1} + K_0^{-1})^{-1}\tilde{\y}\bigg).
	\posteq\posteq
\end{align}
See the Appendix for a derivation. One can easily verify that the above simplifies to the marginal likelihood of Eq.~\eqref{eqn:margLike_singleObs} when $J=1$.
\presec
\section{Inference of the Hierarchical Partition}
\label{sec:partition}
\postsec
In the formulation so far, we have assumed that the hierarchical partition $\A$ is given.  A key question is to infer the partition from the data.  Assume that we have prior $p(\A)$ on the hierarchical partition.  Based on the fact that we can analytically compute $p(Y\mid \A)$, we can use importance sampling or independence chain Metropolis Hastings to draw samples from the posterior $p(\A \mid Y)$.
\prepar
\paragraph{Partition Prior} For the hierarchical partition, we consider a prior solely on the partition points rather than taking tree level into account as well.  Recall that our partition trees hierarchically join neighboring contiguous regions in $\X$.  Because of our time-series analysis focus, we assume $\X \subset \Re$. We define a distribution $F$ on $\X$; the prior probability of $\A$ with partition points $\{z_1,\dots, z_{2^{L-1}-1}\}$ is given by $p(\A) = \prod_i F(z_i)$.  Generatively, one can think of drawing $2^{L-1}-1$ partition points and deterministically forming a balanced binary tree $\A$ from these.  For multidimensional $\X$, one could use Voronoi tessellation and graph matching to build the tree from the randomly selected $z_i$.  Such a prior allows for trivial specification of a uniform distribution on $\A$ (simply taking $F$ uniform on $\X$) or for eliciting prior information on changepoints, such as based on physiological information for the MEG data.  In contrast, eliciting such information in a level-dependent setup is not straightforward.  Also, despite common deployment, a prior that takes the partition point at level $\ell$ as uniformly distributed over the parent set $\A_i^{\ell-1}$ yields high mass on $\A$ with small $A_i^\ell$.  This property is undesirable because it leads to trees with highly unbalanced partitions.

\begin{wrapfigure}{r}{.32\textwidth}
\centering
\vspace{-0.15in}
\includegraphics[width=.3\textwidth]{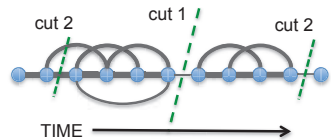}
\vspace*{-8pt}
\caption{\small Illustration of cutpoints dividing contiguous segments at points of low correlation.} 
\label{fig:ncut}
\end{wrapfigure}

Our resulting inferences perform Bayesian model averaging over trees.  As such, even though we specify a prior on partitions with $2^{L-1}-1$ changepoints, the resulting functions can appear to adaptively use fewer by averaging over the uncertainty in the discontinuity location.
%
%
\prepar
\paragraph{Partition Proposal} Although stochastic tree search algorithms tend to be inefficient in general, we can harness the well-defined correlation structure associated with a given hierarchical partition to much more efficiently search the tree space.  One can think of every observed location $x_i$ as a node in a graph with edge weights between $x_i$ and $x_j$ defined by the magnitude of the correlation of $y_i$ and $y_j$.  Based on this interpretation, the partition points of $\A$ correspond to graph cuts that bisect small edge weights, as graphically depicted in Fig.~\ref{fig:ncut}. As such, we seek a method for hierarchically cutting a graph.  Given a cost matrix $W$ with $w(u,v)$ defined for all pairs of nodes $u,v$ in a set $V$, the \emph{normalized cut} metric~\cite{Shi:00} for partitioning $V$ into disjoint sets $A$ and $B$ is given by
\begin{align}
	\mbox{ncut}(A,B) = \mbox{cut}(A,B)\left[\mbox{assoc}(A,V)^{-1} + \mbox{assoc}(B,V)^{-1}\right],
\end{align}
where $\mbox{cut}(A,B) = \sum_{u \in A,v \in B} w(u,v)$ and $\mbox{assoc}(A,V) = \sum_{u\in A,v\in V} w(u,v)$. Typically, the cut point is selected as the minimum of the metric $\mbox{ncut}(A,B)$ computed over all possible subsets $A$ and $B$.  The normalized cut metric balances the connectivity measure between $A$ and $B$ with the connectivity between the elements in $A$ (or $B$) and \emph{all} elements, thus avoiding cuts that separate small sets.  Fig.~\ref{fig:MEGexample} shows an example of applying a greedy normalized cut algorithm to MEG data.

%
Instead of deterministically selecting cut points, we employ the normalized cuts objective as a proposal distribution.  Let the cost matrix $W$ be the absolute value of the empirical correlation matrix computed from replicates $\{\y^{(1)},\dots,\y^{(J)}\}$ (see Fig.~\ref{fig:MEGexample}).  Due to the natural ordering of our locations $x_i \in \X \subset \Re$, the algorithm is straightforwardly implemented.  We step down the hierarchy, first proposing a cut of $\A^0$ into $\{\A^1_1,\A^1_2\}$ with probability 
\begin{align}
q(\{\A^1_1,\A^1_2\}) \propto \mbox{ncut}(\A^1_1,\A^1_2)^{-1}.
\end{align}
At level $\ell$, each $\A_i^\ell$ is partitioned via a normalized cut proposal based on the submatrix of $W$ corresponding to the locations $x_i \in A_i^\ell$.  The probability of any partition $\A$ under the specified proposal distribution is simply computed as the product of the sequence of conditional probabilities of each cut.  This procedure generates cut points only at the observed locations $x_i$.  More formally, the partition point in $\X$ is proposed as uniformly distributed between $x_i$ and $x_{i+1}$.  Extensions to multidimensional $\X$ rely on spectral clustering algorithms based on the graph Laplacian~\cite{Luxburg:07}.
\prepar
\paragraph{Markov Chain Monte Carlo}
An importance sampler draws hierarchical partitions $\A^{(m)} \sim q$, with the proposal distribution $q$ defined as above, and then weights the samples by $p(\A^{(m)})/q(\A^{(m)})$ to obtain posterior draws~\cite{RobertCasella}. Such an approach is naively parallelizable, and thus amenable to efficient computations, though the effective sample size may be low if $q$ does not adequately match the posterior $p(\A \mid Y)$.  Alternatively, a straightforward independence chain Metropolis Hastings algorithm (see Appendix) is defined by iteratively proposing $\A' \sim q$ which is accepted with probability $\min\{r(\A' \mid \A),1\}$ where $\A$ is a previous sample of a hierarchical partition and 
%
\begin{align}
	r(\A' \mid \A) = p(Y \mid A')p(A')q(A)/[p(Y \mid A)p(A)q(A')].
\end{align}
The tailoring of the proposal distribution $q$ to this application based on normalized cuts dramatically aids in improving the acceptance rate relative to more naive tree proposals.  However, the acceptance rate tends to decrease as higher posterior probability partitions $\A$ are discovered, especially for many level trees defined over large spaces $\X$ for which the search space is larger.  

One benefit of the MCMC approach over importance sampling is the ability to include more intricate tree proposals to increase efficiency.  We choose to interleave both local and global tree proposals.  At each iteration, we first randomly select a node in the tree (i.e., a partition set $\A_i^\ell$) and then propose a new sequence of cuts for all children of this node.  When the root node is selected, corresponding to $\A^0$, the proposal is equivalent to the global proposals previously considered.  We adapt the proposal distribution for node selection to encourage more global searches at first and then shift towards a greater balance between local and global searches as the sampling progresses. Sequential Monte Carlo methods~\cite{DelMoral:06} can also be considered, with particles generated as global proposals.
%
%
\prepar
\paragraph{Computational Complexity} The per iteration complexity is $O(n^3)$, equivalent to a typical likelihood evaluation under a GP prior.  Using dynamic programming, the cost associated with the normalized cut proposal is $O(n^2(L-1))$.  Standard techniques for more efficient GP computations are readily applicable, 
as well as extensions that harness the additive block structure of the covariance.
\presec 
\section{Related Work}
\label{sec:related}
\postsec
Various aspects of the mGP have similarities to other models proposed in the literature that primarily fall into two main categories: (i) GPs defined over a partitioned input space, and (ii) collections of GPs defined at tree nodes.  The treed GP~\cite{Gramacy:08} captures non-stationarities by defining independent GPs at the leaves of a Bayesian CART-partitioned input space. 
The related approach of~\cite{Kim:05} assumes a Voronoi tessellation. These methods capture abrupt changes, but do not allow for long-range dependencies nor a functional data hierarchical structure, both inherent to our multiresolution perspective.  Instead, a main motivation is the resulting computational speed-ups of an independently partitioned GP.  A two-level hierarchical GP also aimed at computational efficiency is considered by~\cite{Park:10}.  This GP approximation takes the upper level to be a coarse-scale GP defined over the centroids of the k-means (or known) input partition, and the lower level to be GPs over each partition set with constant mean given by the parent GP at the given centroid.  The partition model of~\cite{Saatci:10} examines online inference of changepoints with GPs modeling the data within each segment.  

\cite{Jones:11} consider covariance functions defined on a phylogenetic tree such that the covariance between function-valued traits depends on both their spatial distance and evolutionary time spanned via a common ancestor.  \cite{Henao:12} additionally model the phylogenetic tree via the coalescent-based Bayesian hierarchical clustering (BHC) of~\cite{Teh:08b}.  Here, the tree defines the strength and structure of sharing between a collection of functions rather than abrupt changes within the function.  The Bayesian rose tree of~\cite{Blundell:10} focuses on BHC with arbitrary branching structure.  An application is considered where internal nodes are GPs and the data result from a mixture of GPs at the leaves.  Such an approach is fundamentally different from the mGP: observations are not necessarily spatially clustered and each node in the tree defines a GP over the entire input space.  Non-tree-structured mixtures of GP experts were considered in~\cite{Rasmussen:02,Meeds:06}.  Alternatively, multiscale processes have a long history (cf.~\cite{Willsky:02}): the variables define a Markov process on a typically balanced, binary tree and higher-level nodes capture coarser level information about the process.  In contrast, the higher level nodes in the mGP share the same temporal resolution and only vary in smoothness.

At a high level, the mGP differs from previous GP-based tree models in that the nodes of our tree represent GPs over a contiguous subset of the input space $\X$ (or induced Gaussian random vectors of varying dimension) constrained in a hierarchical fashion.  Thus, the mGP combines ideas of GP-based tree models and GP-based partition models.

One can formulate an mGP as an additive GP where each GP in the sum decomposes independently over the level-specific partition of the input space $\mathcal{X}$.  The additive GPs of~\cite{Duvenaud:11} instead focus on coping with multivariate inputs, 
in a similar vain to hierarchical kernel learning~\cite{Bach:09}.  Thus, additive GPs address an inherently different task.  Another formulation related in title, but fundamentally different is the hierarchical GP latent variable model of~\cite{Lawrence:07}.  The formulation takes latent variables at nodes of a fixed tree that are related via GP mappings.

Finally, the hierarchical partitions are reminiscent of those associated with P\'{o}lya trees, and clearly density estimation and regression are closely coupled tasks.  However, the standard P\'{o}lya tree assumes a fixed partitioning scheme.  More recently, randomized and unbalanced partitions were considered in the optional Polya tree~\cite{Ma:11}, but with a scheme tailored to the density estimation task.  
%
%
\presec
\section{Results}
\label{sec:results}
\postsec
\vspace{-3pt}
\subsection{Synthetic Experiments}
\label{sec:sim}
\postssec
\vspace{-3pt}
\begin{figure}[t!]
	\centering
	\begin{tabular}{ccccc} 
		\hspace{-0.2in} \includegraphics[height = 0.95in]{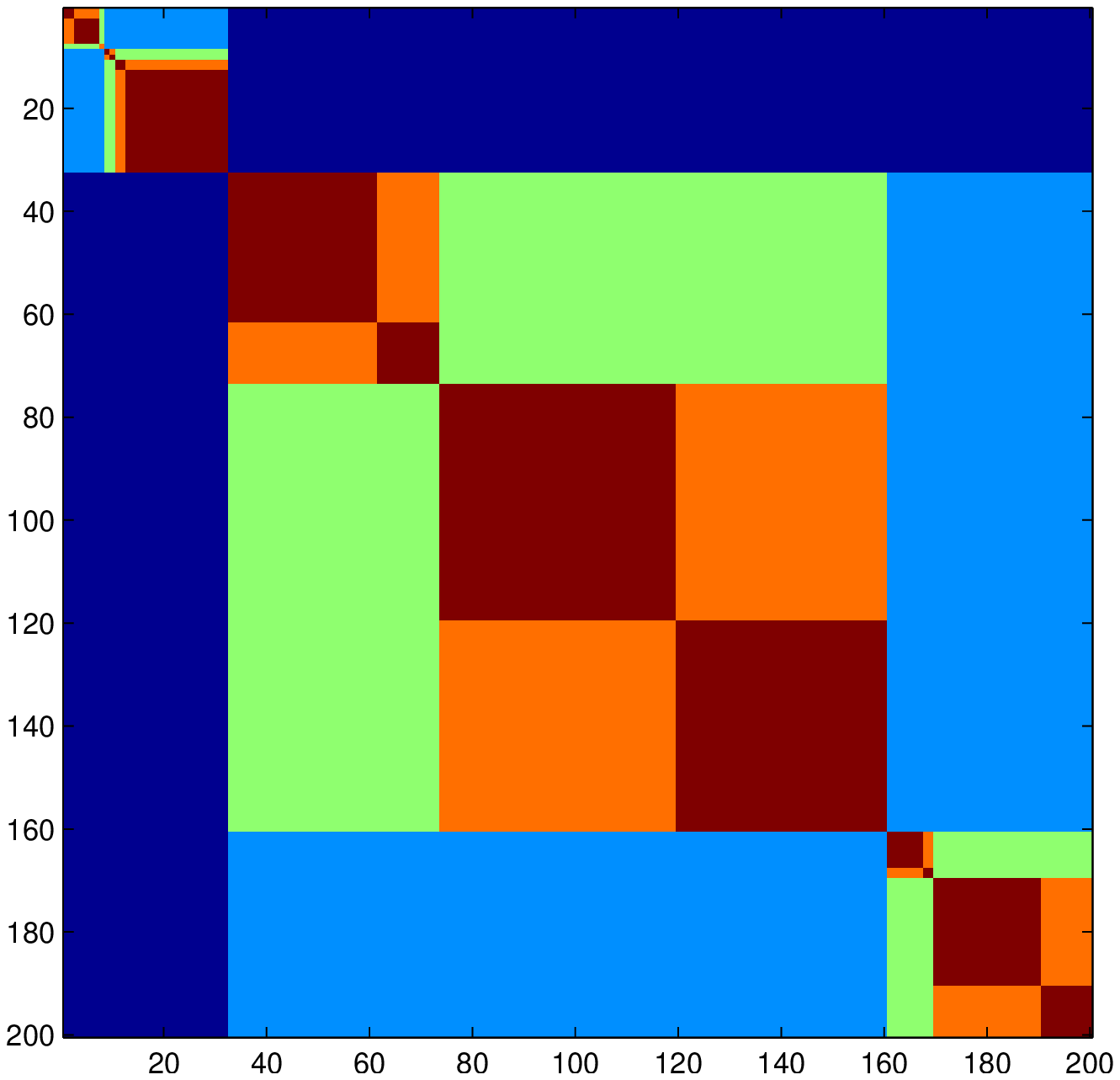} &\hspace{-0.1in} 
		\includegraphics[height = 0.95in]{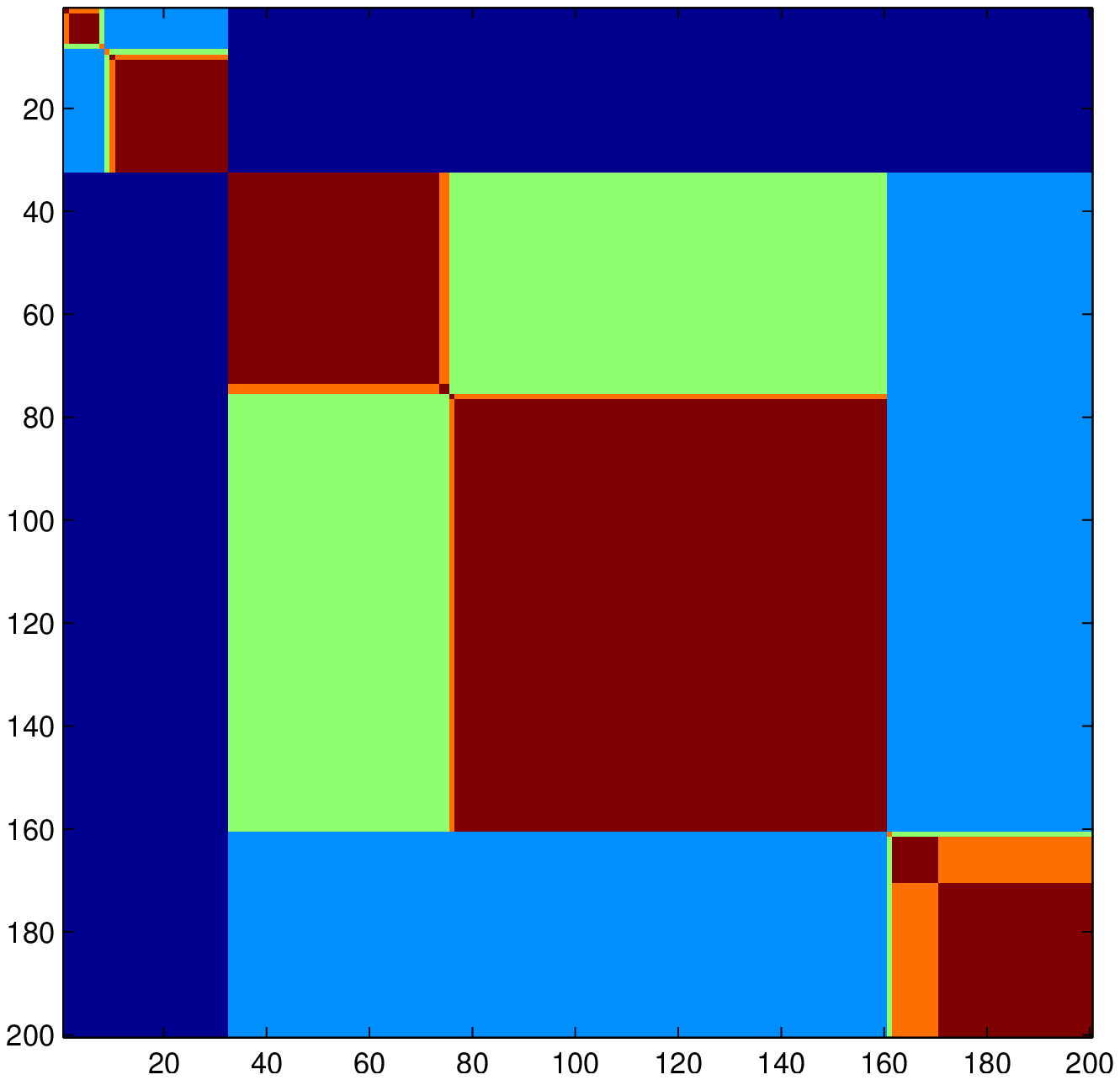} & \hspace{-0.18in}
		\includegraphics[height = 1.1in]{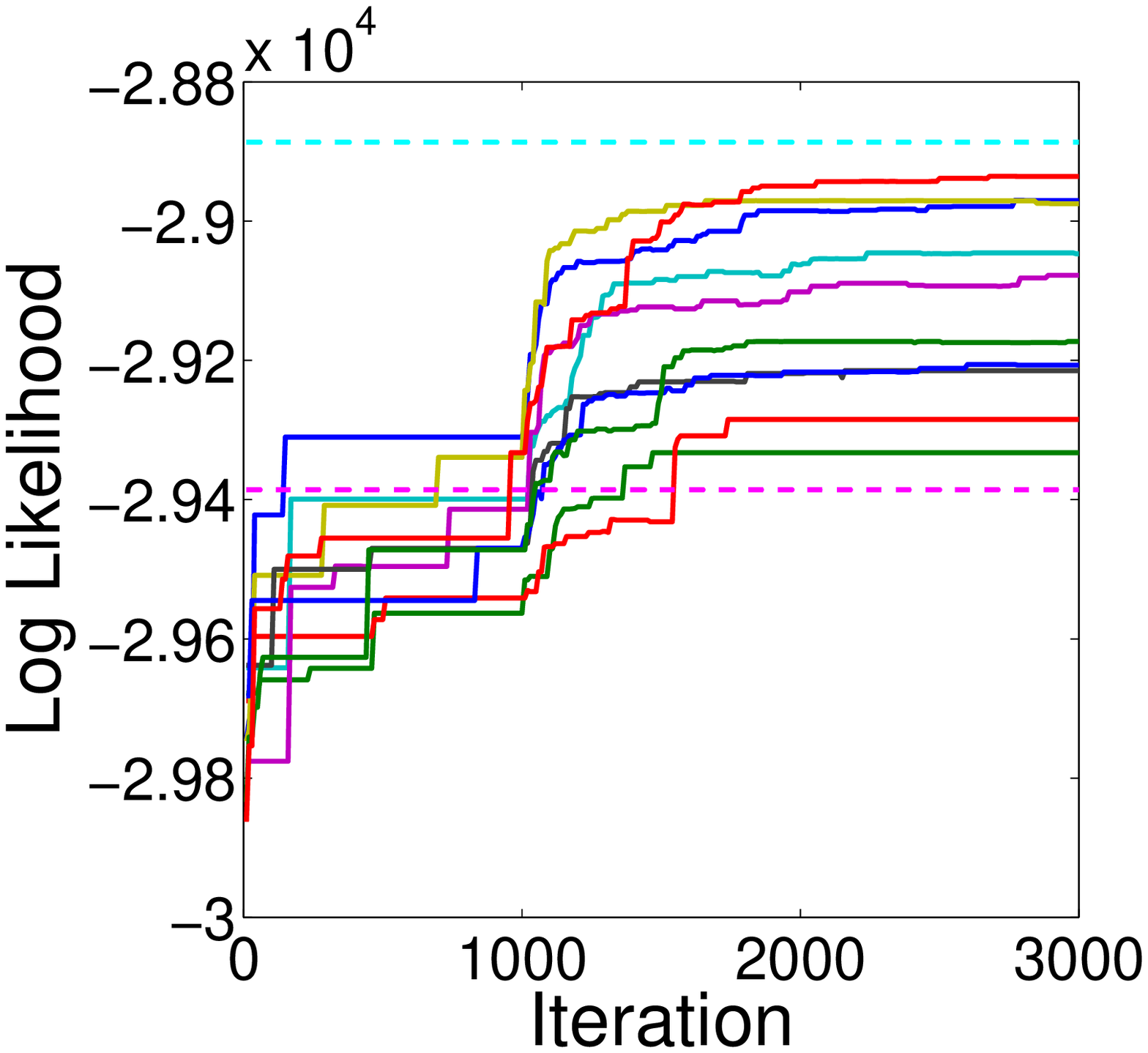} & \hspace{-0.22in}
		\includegraphics[height = 1.05in]{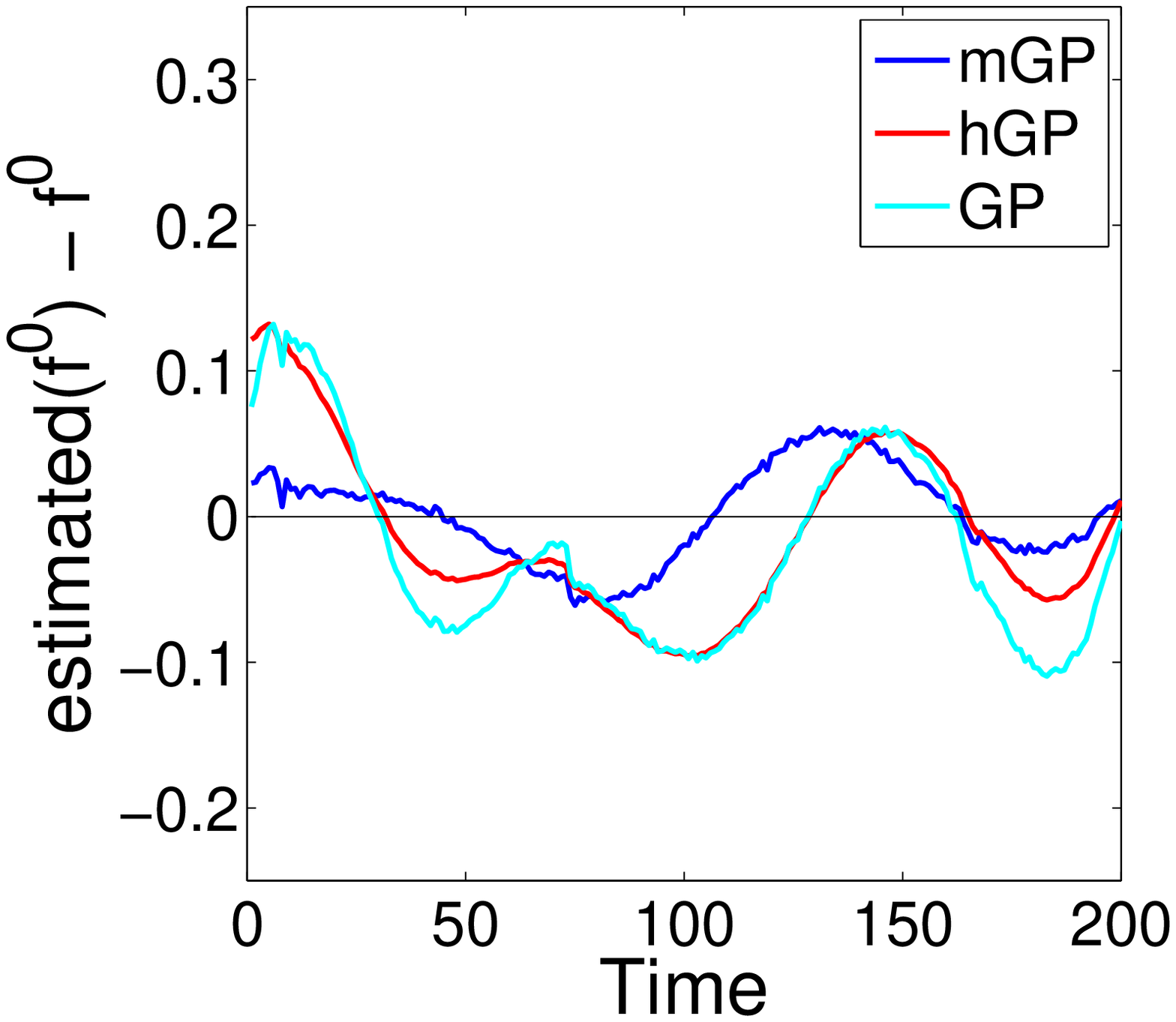} & \hspace{-0.2in} 
		\includegraphics[height = 1.05in]{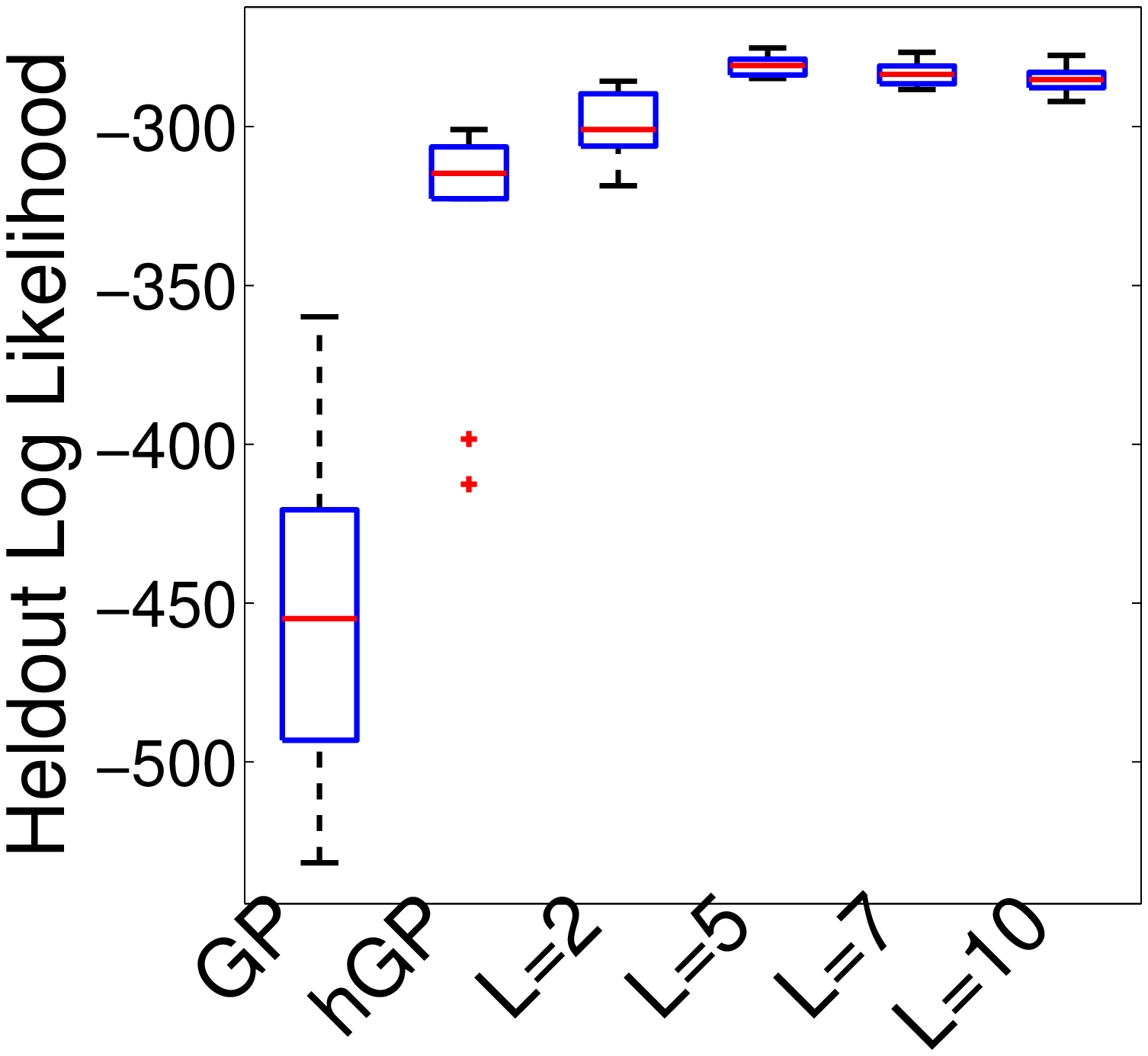}\vspace{-0.05in}\\
		\hspace{-0.2in} {\small (a)} & \hspace{-0.1in} {\small (b)} & \hspace{-0.18in}{\small (c)} & \hspace{-0.22in}{\small (d)}  & \hspace{-0.2in}{\small (e)}\vspace{-0.075in}
	\end{tabular}
	\precap
	\caption{\small For the data of Fig.~\ref{fig:sim}, (a) true and (b) MAP partitions. (c) Trace plots of log likelihood versus MCMC iteration for 10 chains. Log likelihood under the true partition (\emph{cyan}) and minimized normalized cut partition of Fig.~\ref{fig:sim} (\emph{magenta}) are also shown. (d) Errors between posterior mean $f^0$ and true $f^0$ for GP, hGP, and mGP. (e) Predictive log likelihood of 10 heldout sequences for GP, hGP, and mGP with $L=2, 5(true), 7, 10$.} \label{fig:simResults} \postcap \vspace{0.05in}
\end{figure}
To assess our ability to infer a hierarchical partition via the proposed MCMC sampler, we generated 100 replicates of length 200 from a 5-level mGP with a shared parent function $f^0$.  The hyperparameters were set to $\sigma^2 = 0.1$, $\kappa = 10$, $d^\ell = d^0\exp(-0.5(\ell+1))$ for $\ell=0,\dots,L-2$ with $d^0=5$. 
The data are shown in Fig.~\ref{fig:sim}, along with the sample correlation matrix that is used as the cost matrix for the normalized cuts proposals.  

For inference, we set $\sigma^2 = \hat{\sigma}^2/3$ and $d^\ell = (\hat{\sigma}^2/3)\exp(-0.5\ell)$, where $\hat{\sigma}^2$ is the average time-specific sample variance.  
$\kappa$ was as in the simulation.  The hyperparameter mismatch demonstrates some robustness to mispecification.  For a uniform prior $p(\A)$, 10 independent MCMC chains were run for 3000 iterations, thinned by 10.  The first 1000 iterations used pure global tree searches; the sampler was then tempered to uniform node proposals.  The effects of this choice are apparent in the likelihood plot of Fig.~\ref{fig:simResults}, which also displays the true hierarchical partition and MAP estimate.  Compare to the normalized cut partition of Fig.~\ref{fig:sim}, especially at the important level 1 cut. The full simulation study took less than 7 minutes to run on a single 1.8 GHz Intel Core i7 processor.

To assess sensitivity to the choice of $L$, we compare the predictive log-likelihood of 10 heldout test sequences under an mGP with 2, 5, 7, and 10 levels.  As shown in Fig.~\ref{fig:simResults}(e), there is a clear gain going from 2 to 5 levels.  However, overestimating $L$ has minimal influence on predictive likelihood since lower tree levels capture finer details and have less overall effect.  We also compare to a single GP and a 2-level hierarchical GP (hGP) (see Sec.~\ref{sec:MEG}).  For a direct comparison, both use squared exponential kernels. Hyperparameters were set as in the mGP for the top-level GP. The total variance was also matched with the GP taking this as noise and the hGP splitting between level 2 and noise.  In addition to better predictive performance, Fig.~\ref{fig:simResults}(d) shows the mGP's improved estimation of $f^0$.
%
\pressec
\subsection{MEG Analysis}
\label{sec:MEG}
\postssec
We analyzed magnetoencephalography (MEG) recordings of neuronal activity collected from a helmet with gradiometers distributed over 102 locations around the head\footnote{The MEG machine has three sensors at each helmet position: two gradiometers and one magnetometer.  For simplicity, we only consider one gradiometer per location.}.  The gradiometers measure the spatial gradient of the magnetic activity in Teslas per meter (T/m)~\cite{Hansen2010}.  Since the firings of neurons in the brain only induce a weak magnetic field outside of the skull, the signal-to-noise ratio of the MEG data is very low and typically multiple recordings, or \emph{trials}, of a given task are collected.  Our MEG data was recorded while a subject viewed 20 stimuli describing concrete nouns (both the written noun and a representative line drawing), with 20 interleaved trials per word.  These concrete nouns fall into four categories: animals, buildings, food and tools. 

Efficient sharing of information between the single trials is important for tasks such as word classification~\cite{Fyshe:12}.  A key insight of~\cite{Fyshe:12} was the importance of capturing the time-varying correlations between MEG sensors for performing classification.  However, the formulation still necessitates a mean model.  \cite{Fyshe:12} propose a 2-level hierarchical GP (hGP): a parent GP captures the common global trajectory, as in the mGP, and each trial-specific GP is centered about the entire parent function\footnote{The model of~\cite{Fyshe:12} uses an hGP in a latent space.  The mGP could be similarly deployed.}.  This formulation maintains global smoothness at the individual trial level.  The mGP instead models the trial-specific variability with a multi-level tree of GPs defined as deviations from the parent function over local partitions, allowing for abrupt changes relative to the smooth global trajectory. 
\begin{figure}[t!]
	\vspace{-0.0in}
	\begin{center} 
		\begin{tabular}{cccc} 
				\hspace{-0.2in}
				\includegraphics[height = 1in]{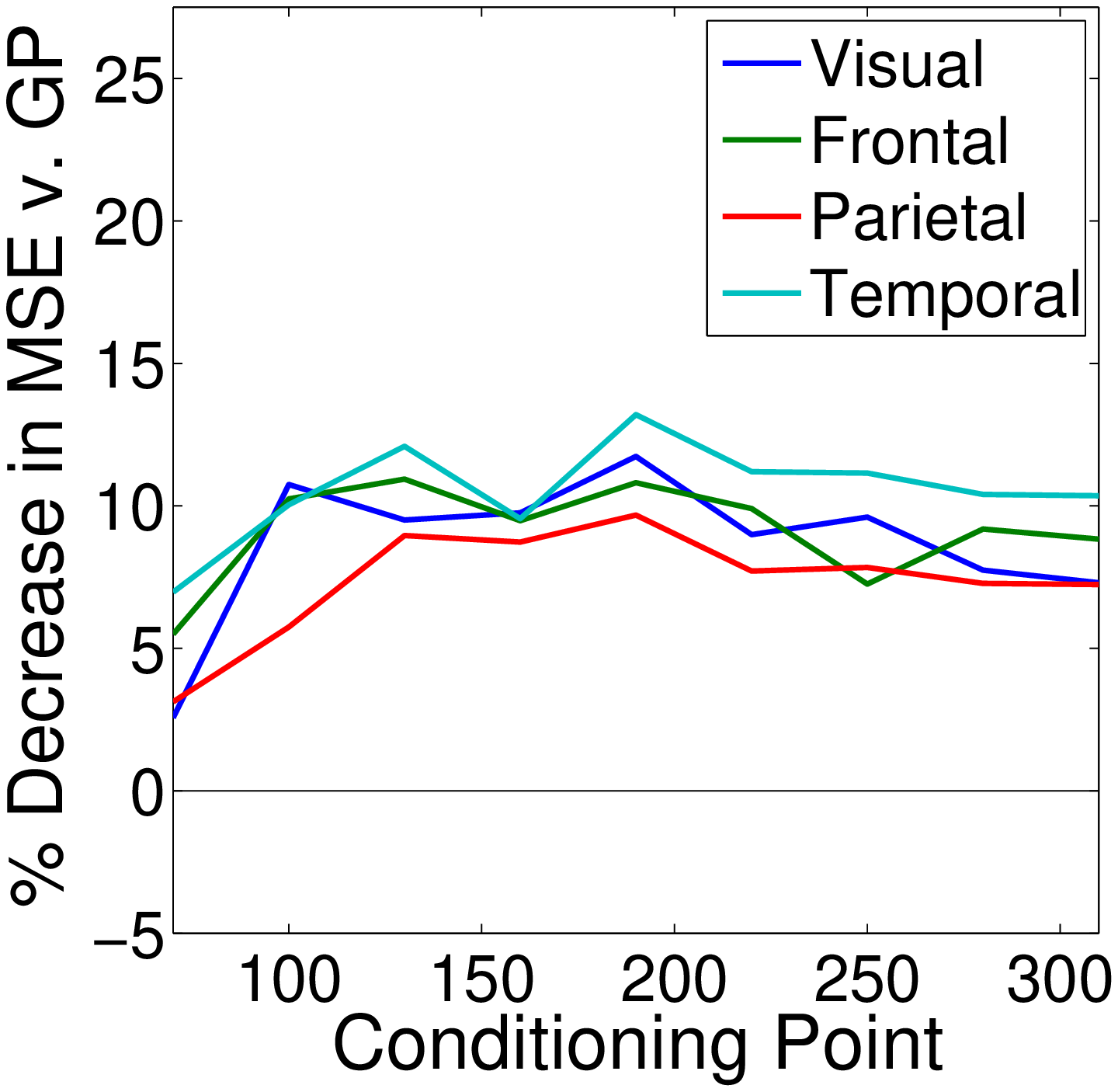} & \hspace{-0.1in}
				\includegraphics[height = 1in]{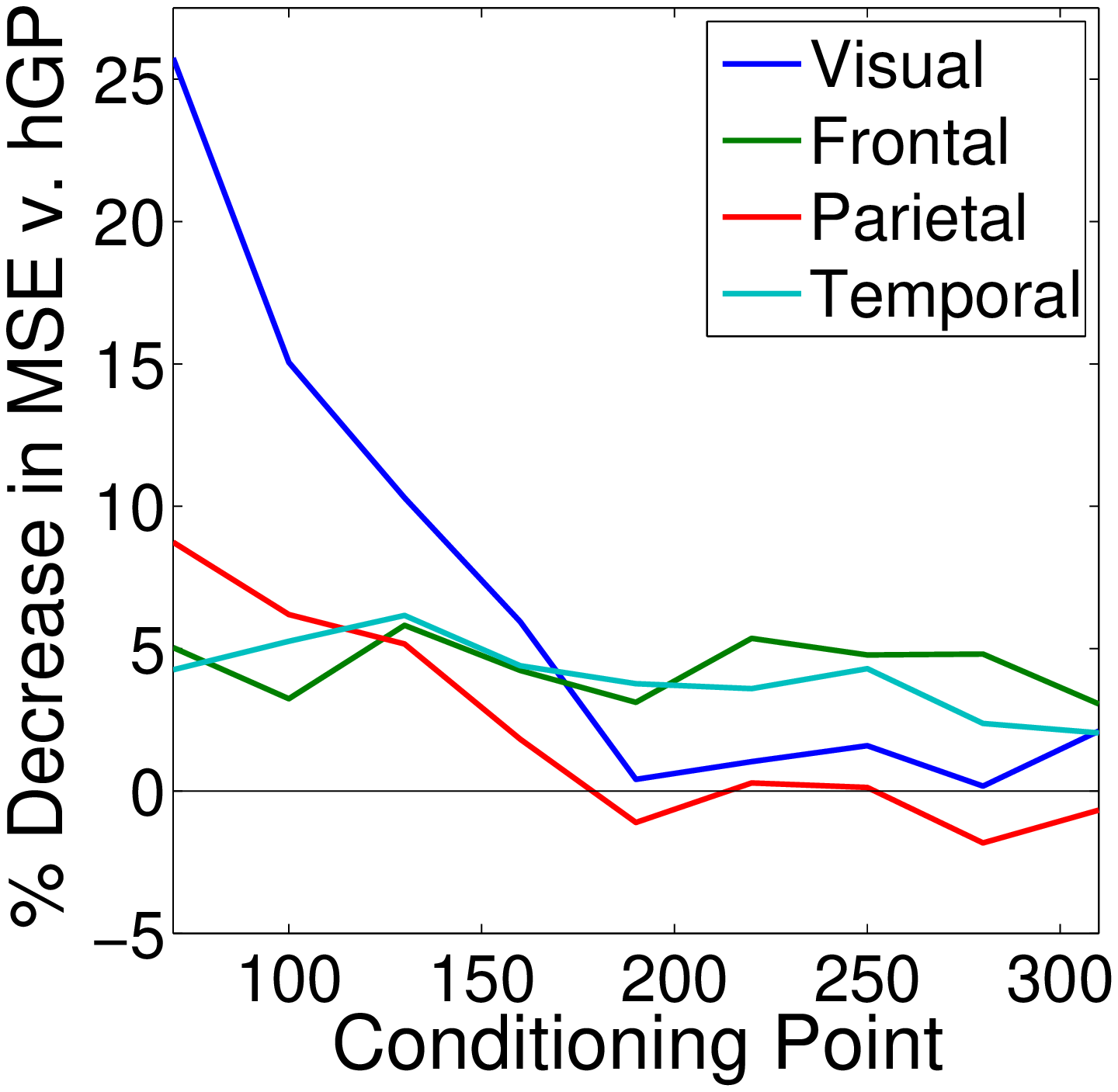} & \hspace{-0.1in}
				\includegraphics[height = 1in]{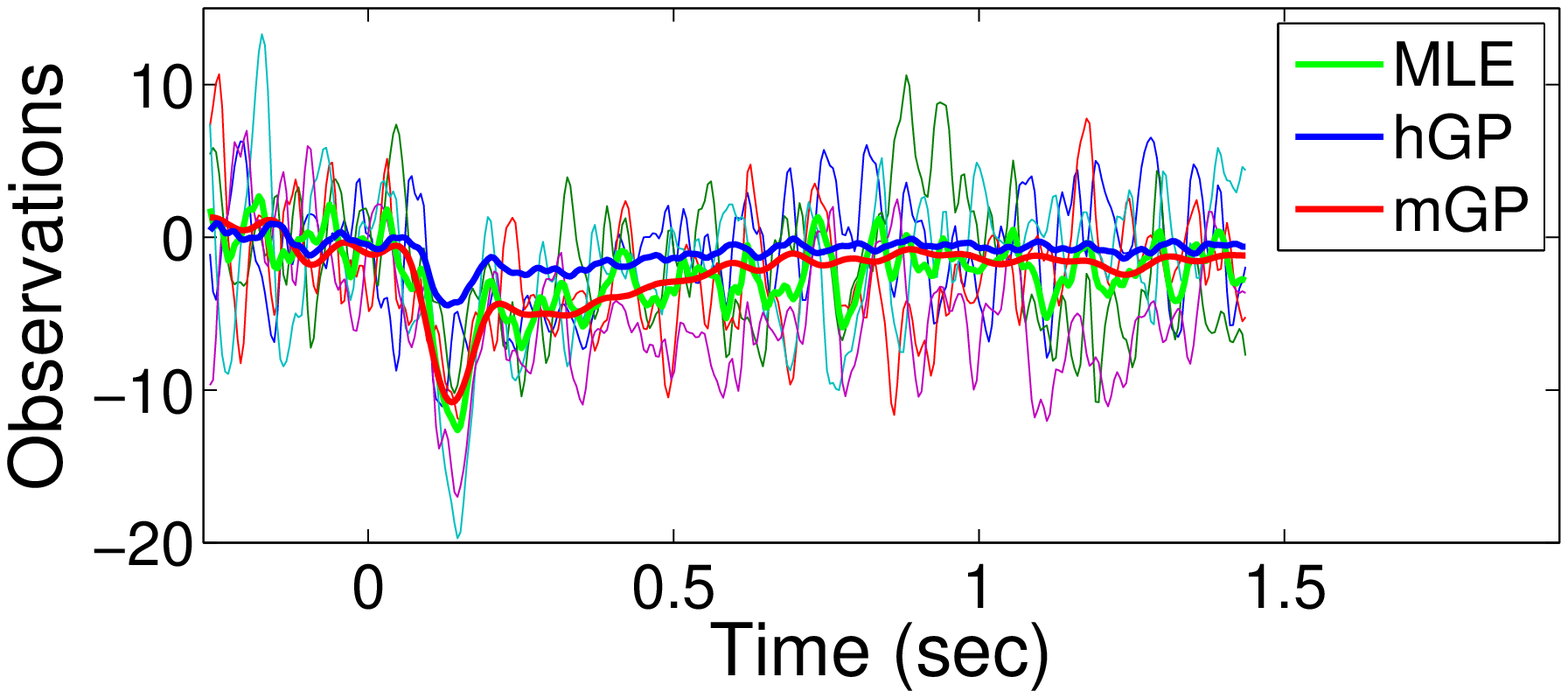} & \hspace{-0.1in}
				\includegraphics[height = 1in]{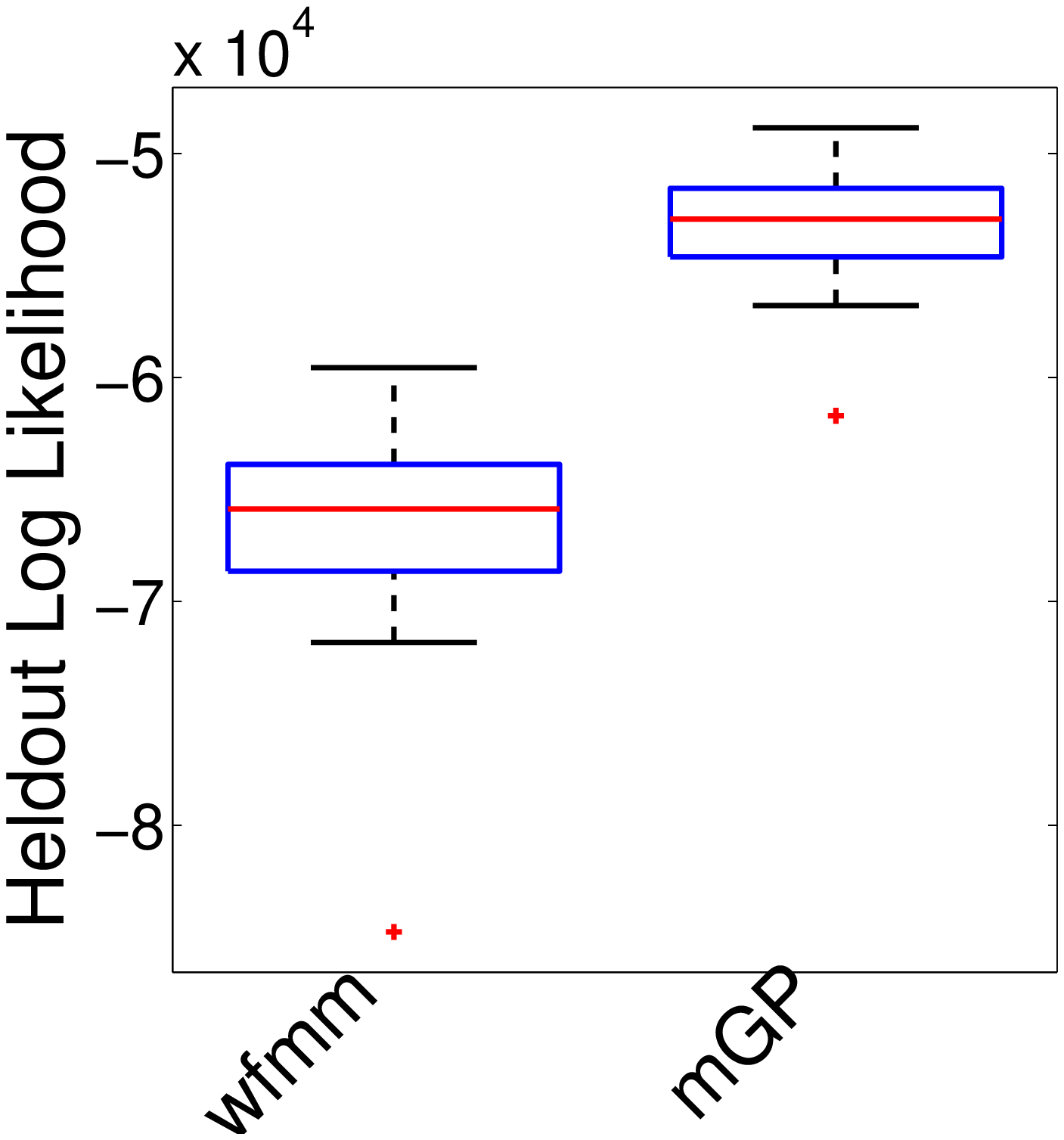}\vspace{-0.06in}\\
				\hspace{-0.2in} {\small (a)} & \hspace{-0.1in} {\small (b)} & \hspace{-0.1in} {\small (c)} & \hspace{-0.1in} {\small (d)}\vspace{-0.05in}
		\end{tabular}
			\precap \vspace{-0.05in}
			\caption{\small Per-lobe comparison of mGP to (a) GP and (b) hGP: For various values of $\tau$, \% decrease in predictive MSE of heldout $y^*_{\tau:\tau+30}$ conditioned on $y^*_{1:\tau-1}$ and 15 training sequences.  (c) For a visual cortex sensor and word \emph{hammer}, plots of test data, empirical mean (MLE), and hGP and mGP predictive mean for entire heldout $\y^*$.  (d) Boxplots of predictive log likelihood of $\y^*$ for the mGP and wavelet-based method of~\cite{Morris:06}. Plots aggregate results over 5 heldout sequences $\y^*$ per word.}
\label{fig:MEGresults}
\postcap \vspace{-0.05in}
	\end{center} 
\end{figure}

For our analyses, we only consider the building (``apartment'', ``barn'', ``church'', ``igloo'', ``house'') and tool (``chisel'', ``hammer'', ``pliers'', ``saw'', ``screwdriver'') words.  Independently for each of the 10 words and 102 sensors, we trained a 5-level mGP using 15 randomly selected trials as training data and the 5 remaining for testing.  Each trial was of length $n=340$.  We ran 3 independent chains of the MCMC sampler for 3000 iterations with both global and local tree searches.  We discarded the first 1000 samples as burn-in and thinned by 10, resulting in 600 hierarchical partition samples $\A^{(m)}$.  The mGP hyperparameters were set exactly as in the simulated study of Sec.~\ref{sec:sim} for structure learning and then optimized over a grid to maximize the marginal likelihood of the training data.  

\begin{wrapfigure}{r}{.4\textwidth}
	\vspace{-0.275in}
\centering
\hspace{-0.1in} 
\begin{center} 
	\begin{tabular}{cc} 
			\hspace{-0.2in} \includegraphics[height = 1in]{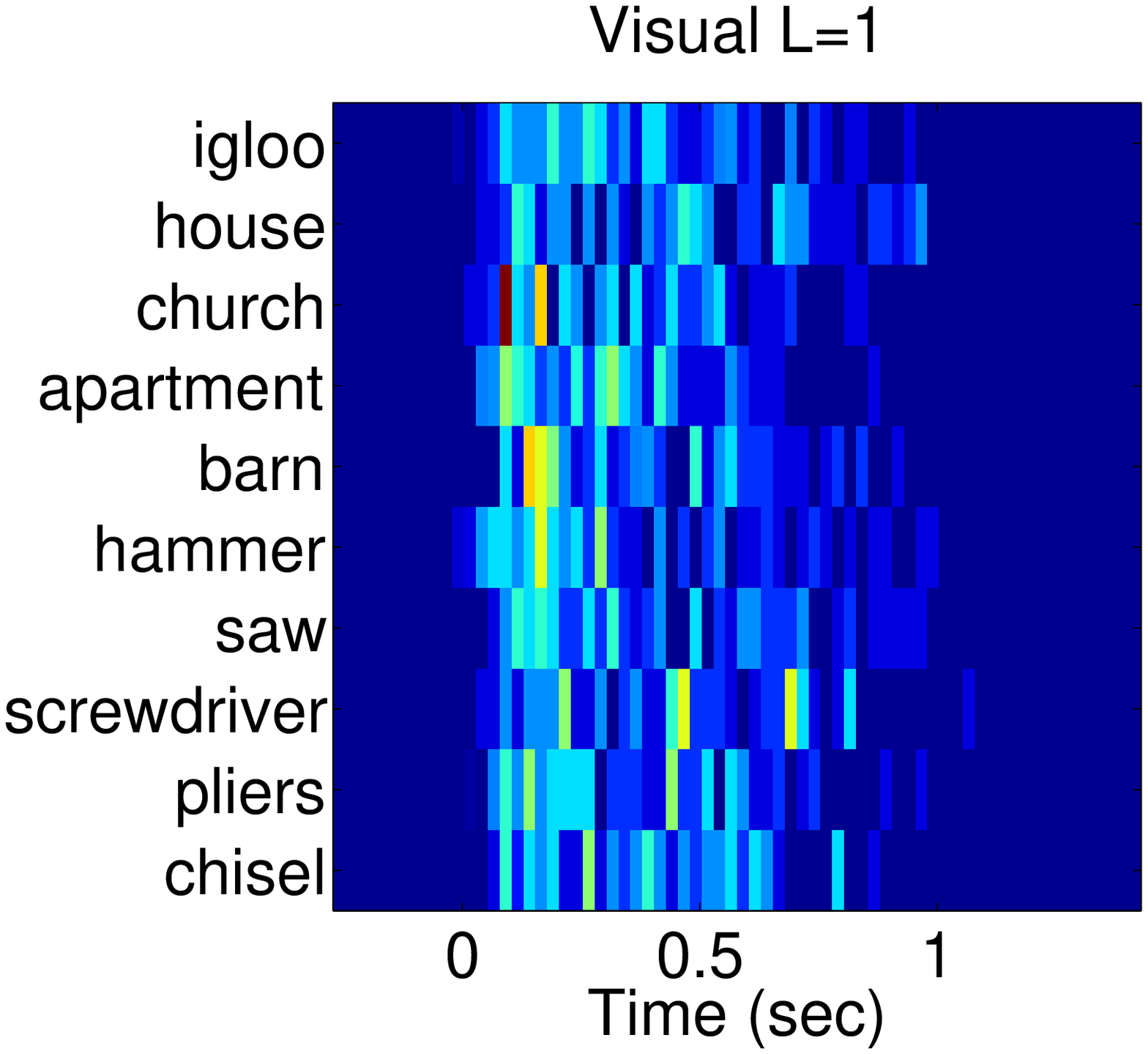} &  
			\hspace{-0.1in} \includegraphics[height = 1in]{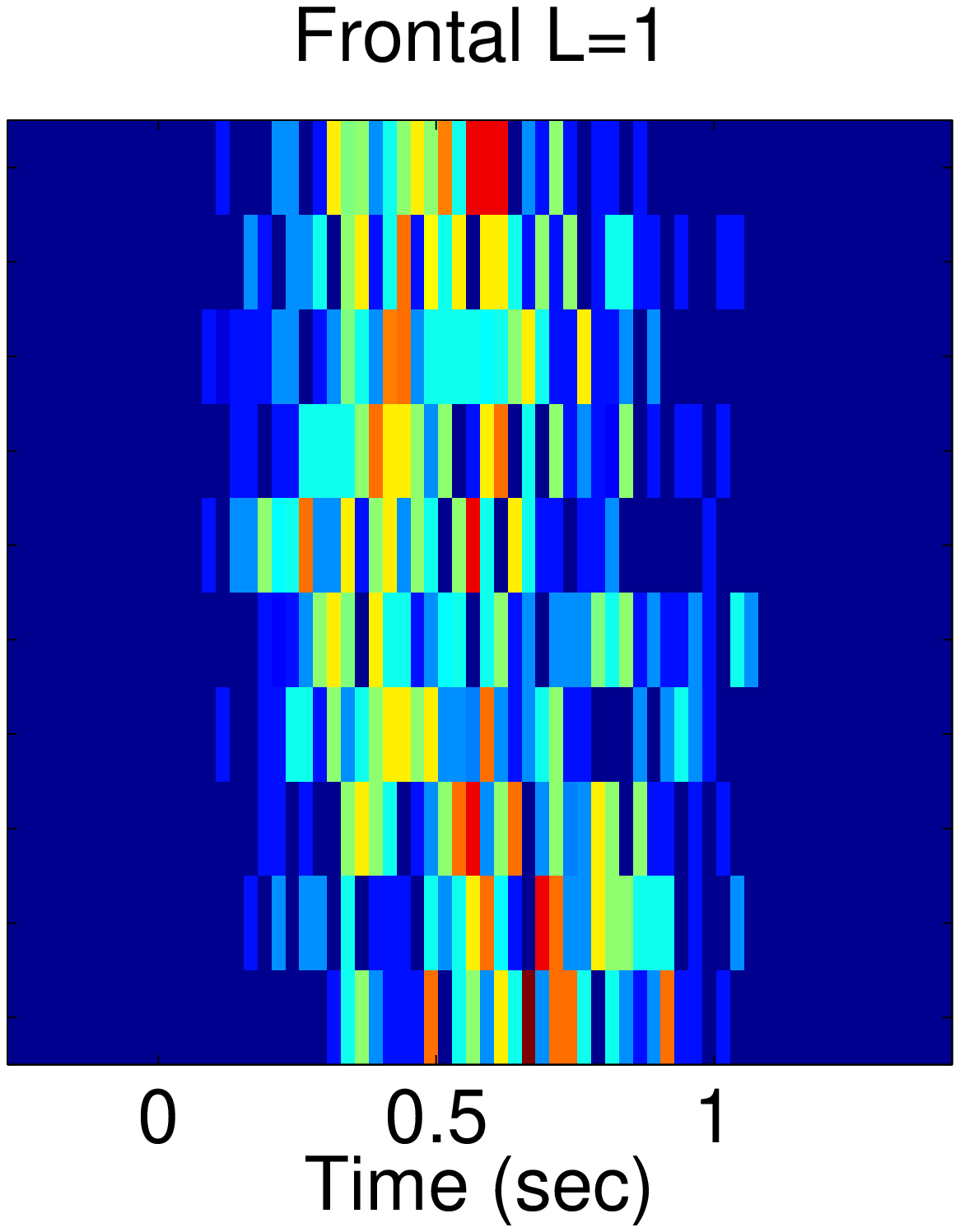}\\
			\hspace{-0.2in} \includegraphics[height = 1in]{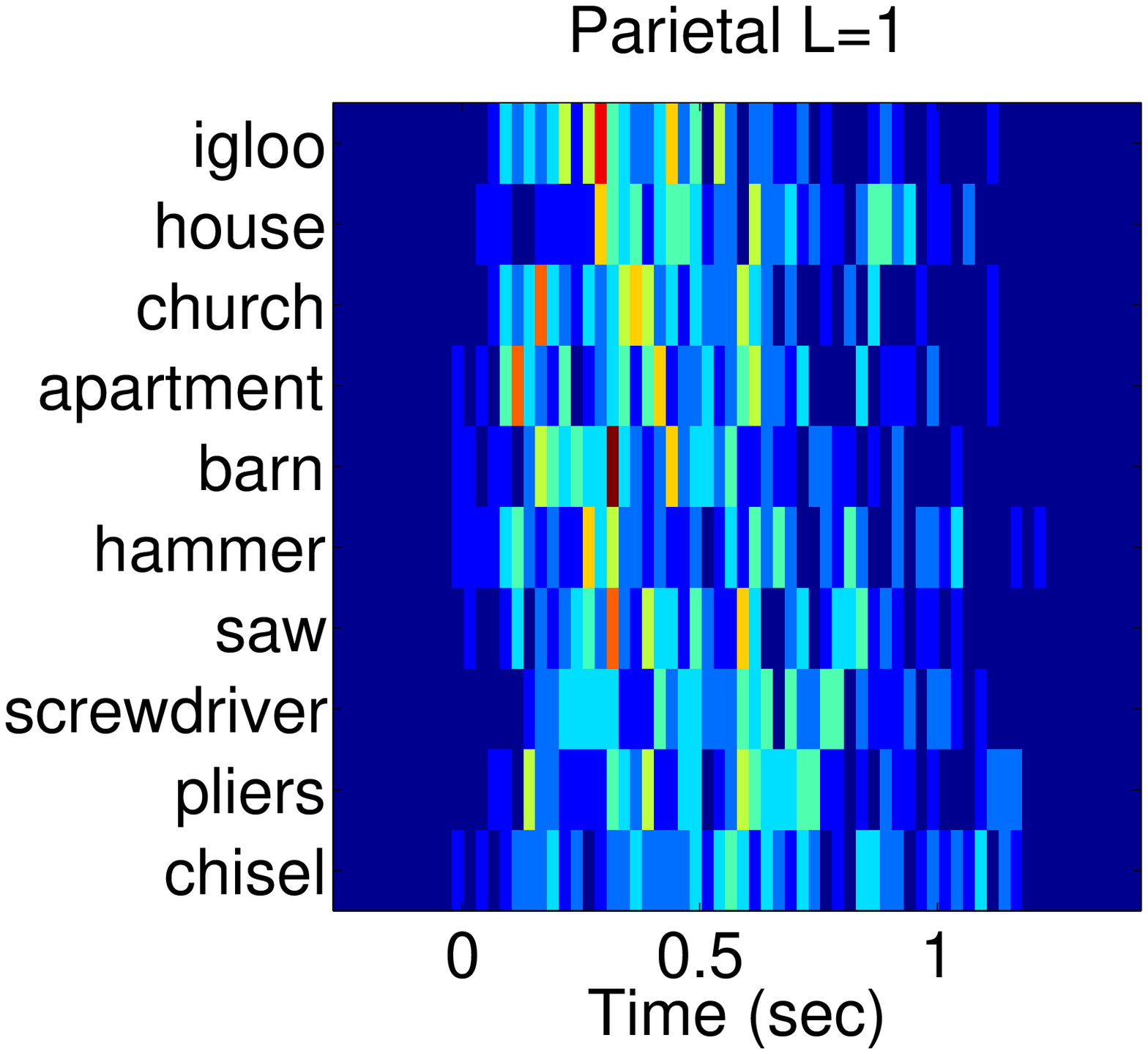} &  
			\hspace{-0.1in} \includegraphics[height = 1in]{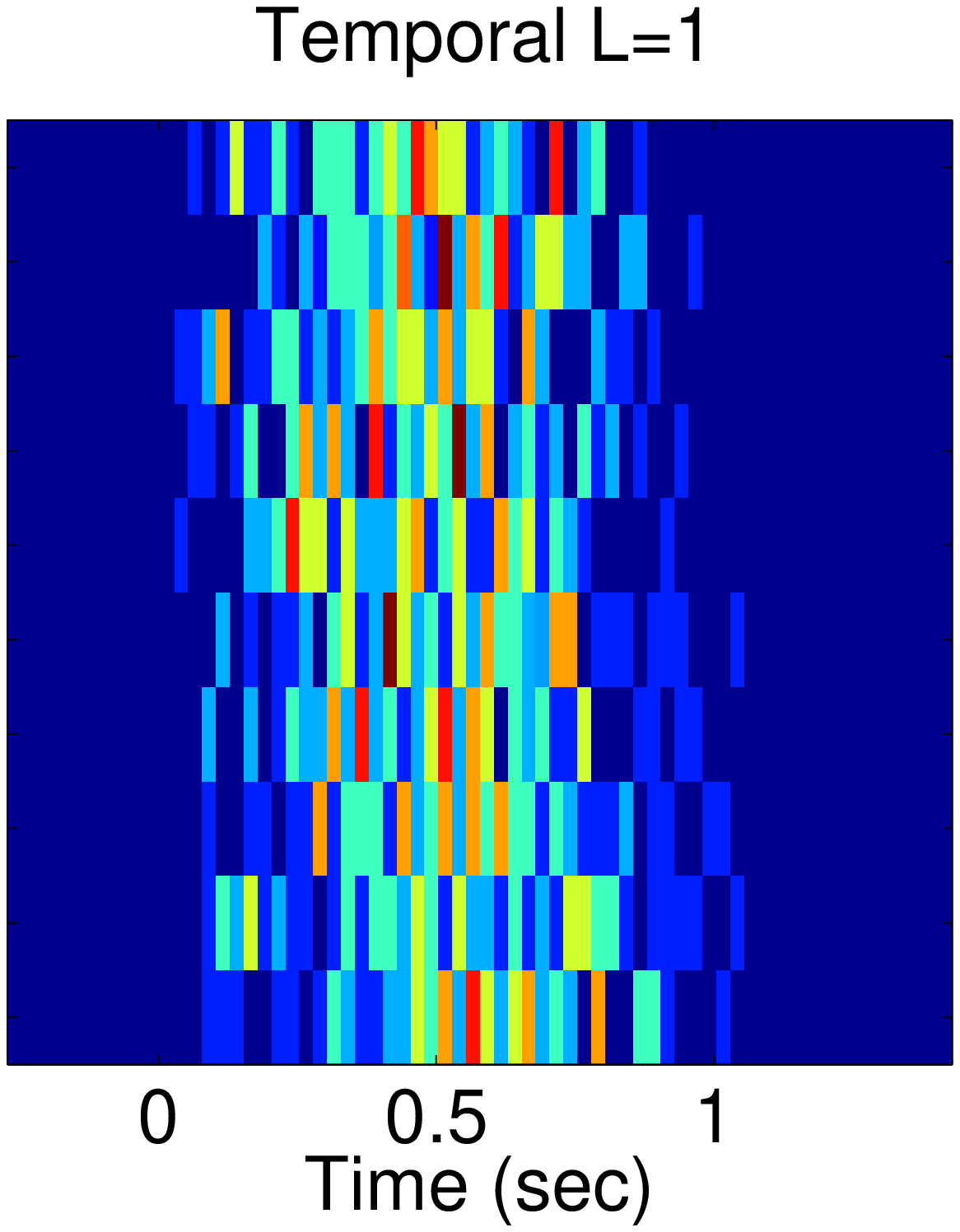}
	\end{tabular}
\end{center}
\vspace*{-10pt}
\precap
\caption{\small Inferred changepoints at level 1 aggregated over sensors within each lobe: visual (\emph{top-left}), frontal (\emph{top-right}), parietal (\emph{bottom-left}), and temporal (\emph{bottom-right}).  
} \label{fig:MEGchangepoints}
\postcap \vspace{0.05in}
\end{wrapfigure}
We compare the predictive performance of the mGP in terms of MSE of heldout segments relative to a GP and hGP, each with similarly optimized hyperparameters.  
The predictive mean conditioned on data up to the heldout time is straightforwardly derived from Eq.~\eqref{eqn:preddist}. For the mGP, the calculation is averaged over the posterior hierarchical partition samples $\A^{(m)}$.  Fig.~\ref{fig:MEGresults} displays the MSEs decomposed by cortical region.  The results clearly indicate that the mGP consistently better captures the features of the data, and particularly for sensors with large abrupt changes such as in the visual cortex.  The heldout replicates for a visual cortex sensor are displayed in Fig.~\ref{fig:MEGresults}(c).  Relative to the hGP, the mGP much better tracks the early jump in activity right after the visual stimulus onset ($t=0$).  The posterior distribution of inferred changepoints at level 1, also broken down by cortical region, are displayed in Fig.~\ref{fig:MEGchangepoints}.  As expected, the visual cortex has the earliest changepoints. 
Similar trends are seen in the parietal lobe that handles perception and sensory integration.  The temporal lobe, which is key in the semantic processing, has changepoints occurring later.  These results concur with the findings of~\cite{Sudre:12}: semantic processing starts between 250 and 600 ms and word length (a visual feature) is decoded most accurately very near the standard 100ms response time (``n100'').

We also compare our predictive performance to that of the wavelet-based functional mixed model (wfmm) of~\cite{Morris:06}.  The wfmm has become a standard approach for functional data analysis since it allows for spiky trajectories and efficient sharing of information between trials.  One limitation, however, is the restriction to a regular grid of observations. Although the wfmm enables analysis in a multivariate setting by incorporating both fixed and random effects, for a direct comparison, we simply apply the wfmm to each word and sensor independently.  Fig.~\ref{fig:MEGresults}(d) shows boxplots of the predictive heldout log likelihood of the test trials under the mGP and wfmm.  In addition to the easier interpretability of the mGP, the predictive performance also exceeds that of the wfmm.
\vspace{-0.03in}
\presec
\section{Discussion}
\label{sec:discussion}
\postsec
\vspace{-0.07in}
The mGP provides a generative framework for characterizing the dependence structure of real data, such as the examined MEG recordings, capturing certain features more accurately than previous models.  In particular, the mGP provides a hierarchical functional data analysis framework for modeling (i) strong, locally smooth sharing of information, (ii) global long-range correlations, and (iii) abrupt changes. The simplicity of the mGP formulation enables further theoretical analysis, for example, combining posterior consistency results from changepoint analysis with those for GPs. Although we focused on univariate time series analysis, our formulation is amenable to multivariate functional data analysis extensions: one can naturally accommodate hierarchical dependence structures through partial sharing of parents in the tree, or possibly via mGP factor models.  Another interesting extension is to incorporate the mGP within a functional ANOVA framework~\cite{Kaufman:10}.

There are many interesting questions relating to the proposed covariance function.  Our fractal specification represents a particular choice to avoid over-parameterization, although alternatives could be considered.  For hyperparameter inference, we anticipate that joint sampling with the partition would mix poorly, and consider it a topic for future exploration. We believe that the proposed mGP represents a powerful, broadly applicable new framework for non-stationary analyses, especially in a functional data analysis setting, and sets the foundation for many interesting possible extensions.

{\small
\section*{Acknowledgements}
\vspace{-0.1in}
The authors would like to thank Alona Fyshe, Gustavo Sudre and Tom Mitchell for their help with data acquisition, preprocessing,
and useful suggestions.  This research was partially supported by the AFOSR under Grant FA9550-10-1-0501 and the National Institute of Environmental Health Sciences (NIEHS) of the NIH under Grant R01 ES017240.
}

\appendix

\section{Derivation of Marginal Conditional Likelihood}

We derive the marginal likelihood as follows.  Throughout, we use $f^0(\x)$ to compactly denote $f^0(x_{1:n})$.  As discussed in the paper, each trial can be described as an independent draw
\begin{align}
	\y^{(j)} \mid f^{0}(\x),\A \sim N\bigg(\y^{(j)};f^{0}(\x),\Sigma \bigg),
\end{align}
where $\Sigma = \sigma^2I_n+\sum_{\ell=1}^{L-1} K_\ell$ and $f^0(\x) \sim N(0,K_0)$.  Therefore, the joint distribution of $Y = \{\y^{(1)},\dots,\y^{(J)}\}$ is given by:
\begin{align}
	p(Y \mid f^0(\x),\A)p(f^0(\x))&= c_1^J c_2 \exp\left(-\frac{1}{2} \sum_i (\y^{(i)} - f^0(\x))'\Sigma^{-1}(\y^{(i)} - f^0(\x)) - \frac{1}{2} \f^{0'}K_0^{-1}f^0(\x)\right)\nonumber\\
	&\hspace{-1in}= c_1^J c_2 \exp\left(-\frac{1}{2} \sum_i \y^{(i)'}\Sigma^{-1}\y^{(i)} +  \f^{0'}\Sigma^{-1}\sum_i \y^{(i)} - \frac{1}{2}\f^{0'}(J\Sigma^{-1} + K_0^{-1})f^0(\x)\right)\nonumber\\
	&\hspace{-1in}= c_1^J c_2 \exp\left(-\frac{1}{2} \sum_i \y^{(i)'}\Sigma^{-1}\y^{(i)} - \frac{1}{2}(f^0(\x)-\phi)'(J\Sigma^{-1} + K_0^{-1})(f^0(\x)-\phi) + \frac{1}{2}\phi' (J\Sigma^{-1} + K_0^{-1})\phi\right),
\end{align}
where $c_1 = ((2\pi)^{n/2}|\Sigma|^{1/2})^{-1}$, $c_2 = ((2\pi)^{n/2}|K_0|^{1/2})^{-1}$, and $\phi = (J\Sigma^{-1} + K_0^{-1})^{-1}\Sigma^{-1}\sum_i \y^{(i)}$.  Then, $p(Y\mid \A) = \int p(Y \mid f^0(\x),\A)p(f^0(\x)) df^0(\x)$ is derived as
\begin{align}
	p(Y\mid \A) &= c_1^J c_2 \exp\left(-\frac{1}{2} \sum_i \y^{(i)'}\Sigma^{-1}\y^{(i)} + \frac{1}{2}\phi' (J\Sigma^{-1} + K_0^{-1})\phi\right)\nonumber\\
	&\hspace{0.5in} \int \exp\left(-\frac{1}{2}(f^0(\x)-\phi)'(J\Sigma^{-1} + K_0^{-1})(f^0(\x)-\phi)\right)df^0(x_{1:N})\\
	&= c_1^J c_2 \exp\left(-\frac{1}{2} \sum_i \y^{(i)'}\Sigma^{-1}\y^{(i)} + \frac{1}{2}\phi' (J\Sigma^{-1} + K_0^{-1})\phi\right)\cdot (2\pi)^{n/2} |J\Sigma^{-1} + K_0^{-1}|^{-1/2}\\
	&= (2\pi)^{-nJ/2}|K_0|^{-1/2}|\Sigma|^{-J/2} |J\Sigma^{-1} + K_0^{-1}|^{-1/2} \nonumber\\
	&\hspace{0.5in}\exp\left(-\frac{1}{2} \sum_i \y^{(i)'}\Sigma^{-1}\y^{(i)} + \frac{1}{2}\phi' (J\Sigma^{-1} + K_0^{-1})\phi\right).
\end{align}

\section{Details on MEG Experiments}

For the MEG experiments, the data from each word $w$ and sensor $p$ were treated independently.  That is, each assumed a unique partition structure for the mGP.  Additionally, for all models the hyperparameters were set in a training-data-driven fashion.

\subsection{Hyperparameter Optimization}

The following describes how the hyperparameter optimization is performed for the MEG comparisons.  In all scenarios, the input space $\X = [1:340]$ was first normalized to take values in $[0:1]$, as was the case for the simulated study.

\paragraph{Gaussian Process} The GP was specified as follows.  The covariance function was taken to be a squared exponential, which for word $w$ and sensor $p$ took the form $c_{w,p} = d_{w,p}\exp(-\kappa ||x-x'||_2^2)$.  The scale parameter was constrained to be a fixed linear function of the average time-specific sample variance $\hat{\sigma}_{w,p}^2$ of the training data: $d_{w,p} = \alpha^0\hat{\sigma}_{w,p}^2$.  Likewise, the nugget noise was of the form $\sigma_{w,p}^2 = \beta\hat{\sigma}_{w,p}^2$.  The parameters $\kappa$, $\alpha^0$, and $\beta$ were optimized on a grid to maximize the marginal likelihood of the training data over all words and sensors.

\paragraph{Hierarchical GP} For the 2-level hierarchical GP (hGP), a squared exponential kernel was also assumed for both levels.  As in~\cite{Fyshe:12}, a single bandwidth parameter was assumed.  In particular, for the shared top level GP, $c^0_{w,p} = d^0_{w,p}\exp(-\kappa ||x-x'||_2^2)$, and for the trial-specific level, $c^1_{w,p} = d^1_{w,p}\exp(-\kappa ||x-x'||_2^2)$.  As in the GP, the hyperparameters were constrained as a function of $\hat{\sigma}_{w,p}^2$: $d^0_{w,p} = \alpha^0\hat{\sigma}_{w,p}^2$, $d^1_{w,p} = \alpha^1\hat{\sigma}_{w,p}^2$, and $\sigma_{w,p}^2 = \beta \hat{\sigma}_{w,p}^2$.  Here, $\kappa$, $\alpha^0$, $\alpha^1$, and $\beta$ were jointly optimized to maximize marginal likelihood.  For numerical reasons, the minimum allowable nugget noise was set to 1\% of  $\hat{\sigma}_{w,p}^2$ (i.e., $\beta = 0.01$).

\paragraph{Multiresolution GP} For the multiresolution GP (mGP), the covariance function was constrained in a similar manner to the simulation study.  In particular, a shared bandwidth $\kappa$ was assumed.  The scale parameters $\{d^0,d^1,\dots,d^{L-1}\}$ were constrained as follows.  The global parent GP was assigned scale $d^0 = \alpha^0\hat{\sigma}_{w,p}^2$.  The scale parameters of the $L-1$ trial-specific levels of the mGP hierarchy were constrained by a fixed functional form as in the simulated data setup, determined by two parameters.  In particular, $d^{\ell} = [\alpha^1\exp(-\rho*\ell)]\hat{\sigma}_{w,p}^2$.  Finally, the nugget noise followed $\sigma_{w,p}^2 = \beta\hat{\sigma}_{w,p}^2$.  In this scenario, $\kappa$, $\alpha^0$, $\alpha^1$, $\rho$, and $\beta$ were jointly optimized to maximize marginal likelihood based on initial samples of tree partitions $\A^{(m)}$ using the hyperparameter settings of the simulated data example.  Again, for numerical reasons, the minimum allowable nugget noise was set to 1\% of  $\hat{\sigma}_{w,p}^2$. 

The optimized hyperparameter values were as follows:
\begin{table}[h]
	\begin{center}
	\begin{tabular}{l|ccccc}
		   & $\kappa$ & $\alpha^0$ & $\alpha^1$ & $\beta$ & $\rho$\\
		\hline
		GP & 1350 & 0.15 & -- & 1 & --\\
		hGP & 13000 & 0.033 & 1 & 0.01 & --\\
		mGP & 900 & 0.1 & 1.67 & 0.01 & 1.1
	\end{tabular}
\end{center}
\end{table}

For the hGP, a large bandwidth (low temporal correlation) is taken to account for the abrupt changes.  Also note that the parent GP was given little variance and instead the variance was attributed to the lower level GP to account for the significant trial-to-trial variation.  The GP accounts for both the large trial-to-trial variability and the abrupt changes through a large nugget noise.  The mGP variance dropped off fairly rapidly with tree level, as indicated by $\rho$.  Note that both the hGP and mGP are able to account for trial-to-trial variability in the tree hierarchy instead of through the nugget noise, as indicated by low values of $\beta$.

In our optimization procedure, we found that the performance of the GP was the most sensitive to changes in the hyperparameter specification.  Both the hGP and GP were fairly robust over a reasonably large range of settings (partially indicated by a flat marginal likelihood of the training data over the range.)

\subsection{MEG Prior Settings}

The hierarchical partition prior $p(\A)$, determined by $F$ on $\X$ as described in Sec. 5, was set as follows for a given word $w$ and sensor $p$.  All of the training data associated with sensor $p$ except for that of the considered word was used to produce a recursively minimized normalized cut partition for a 4-level tree.  The associated strength of each cut (i.e., amount of empirical correlation cut) was also recorded.  $F$ was then defined as a kernel-smoothed version of the cut points and associated cut strengths, along with baseline mass at all points.  This prior was used to mimic the information that might be garnered from a domain expert.  Experiments were also run under a uniform prior and produced nearly identical results after burn-in.  The aggregated posterior changepoints samples, depicted in Fig. 8 only for level 1, were clearly different from the prior setting, demonstrating learning of the partition points (not dominated by the prior setting).

\subsection{Additional Figures}

We provide some additional figures related to the MEG results presented in Fig. 7 of the main paper.

\begin{figure}[t!]
	\vspace{-0.0in}
	\begin{center} 
		\begin{tabular}{ccc} 
				\hspace{-0.2in}
				\includegraphics[height = 1.2in]{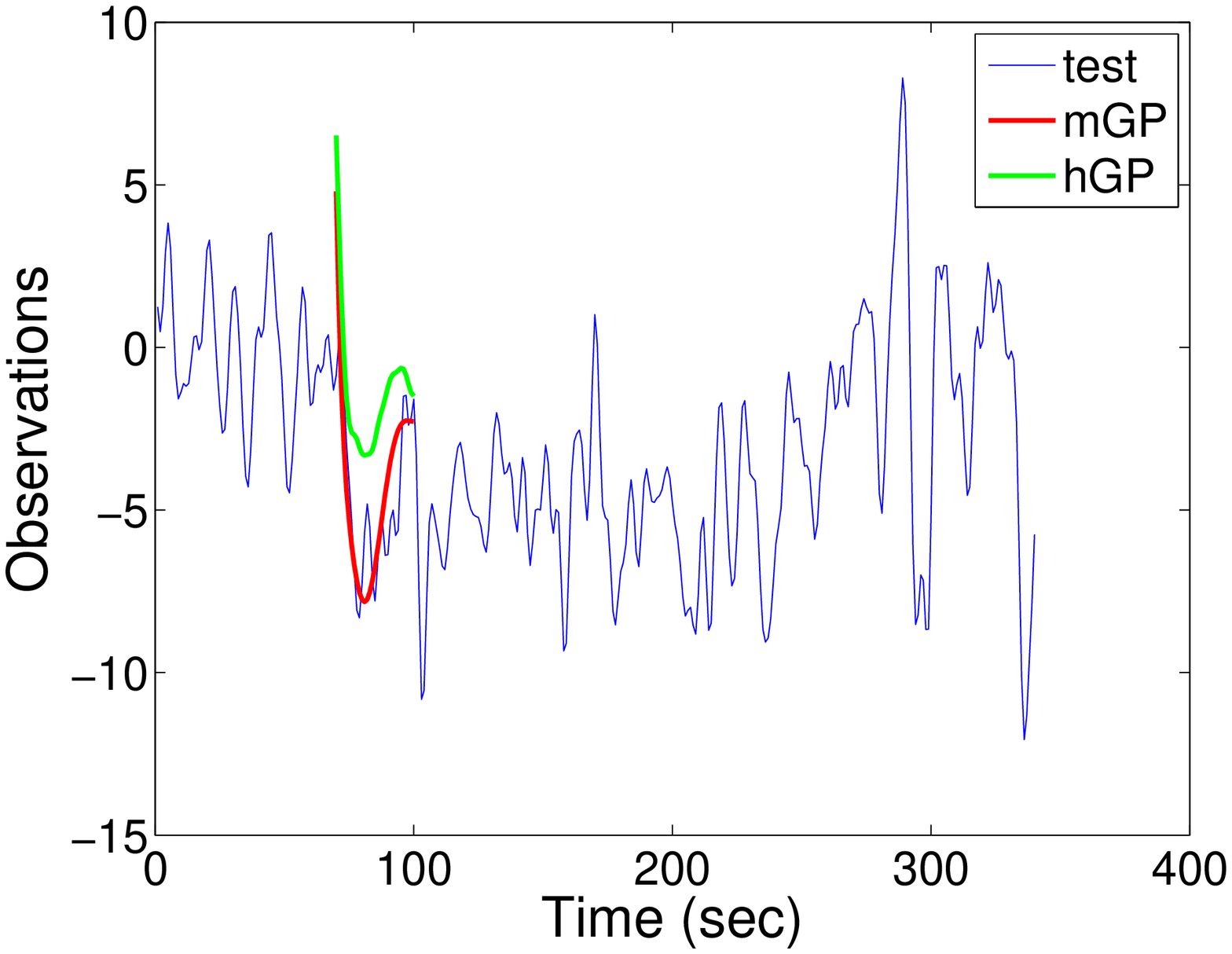} & \hspace{-0.1in}
				\includegraphics[height = 1.2in]{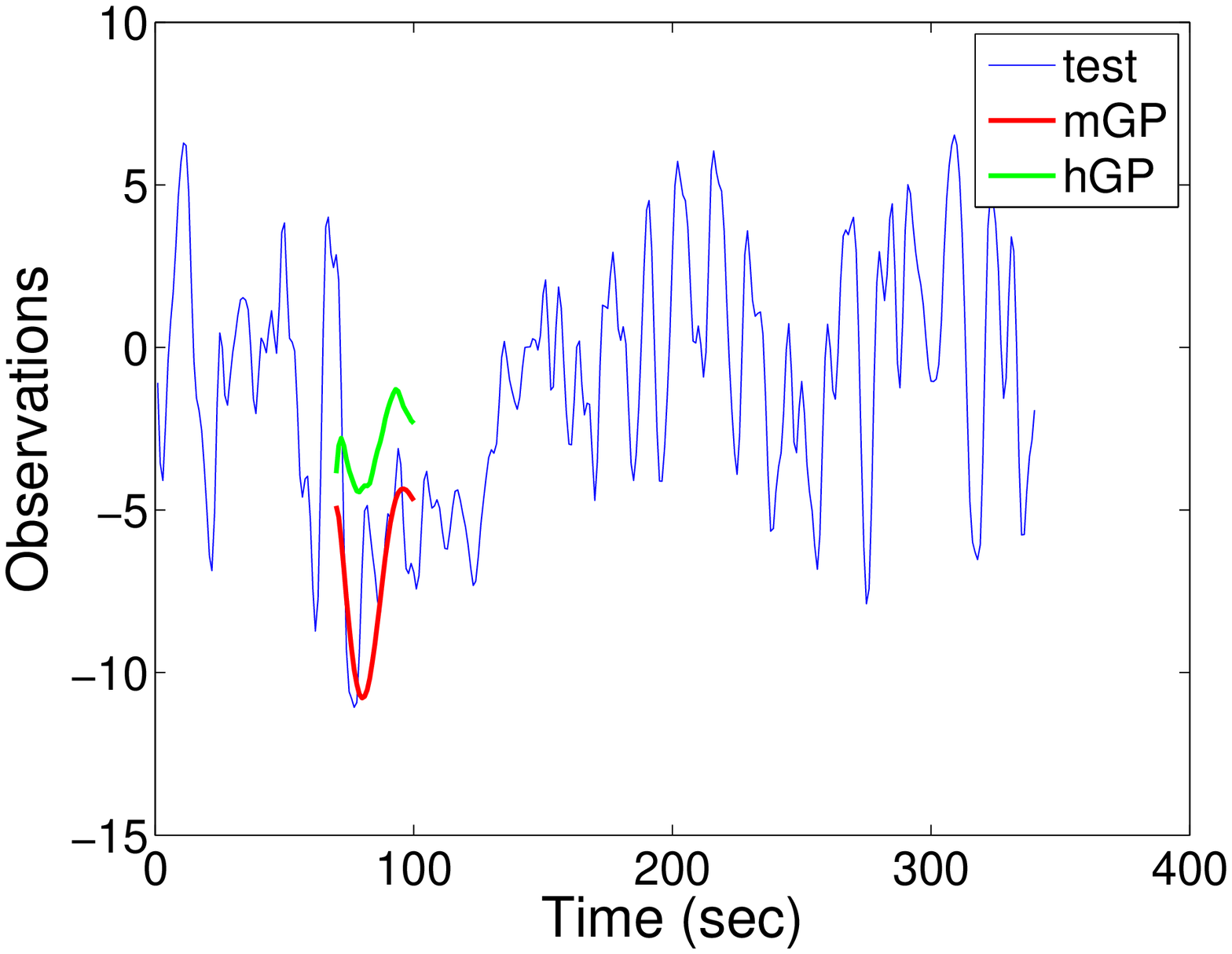} & \hspace{-0.1in}
				\includegraphics[height = 1.2in]{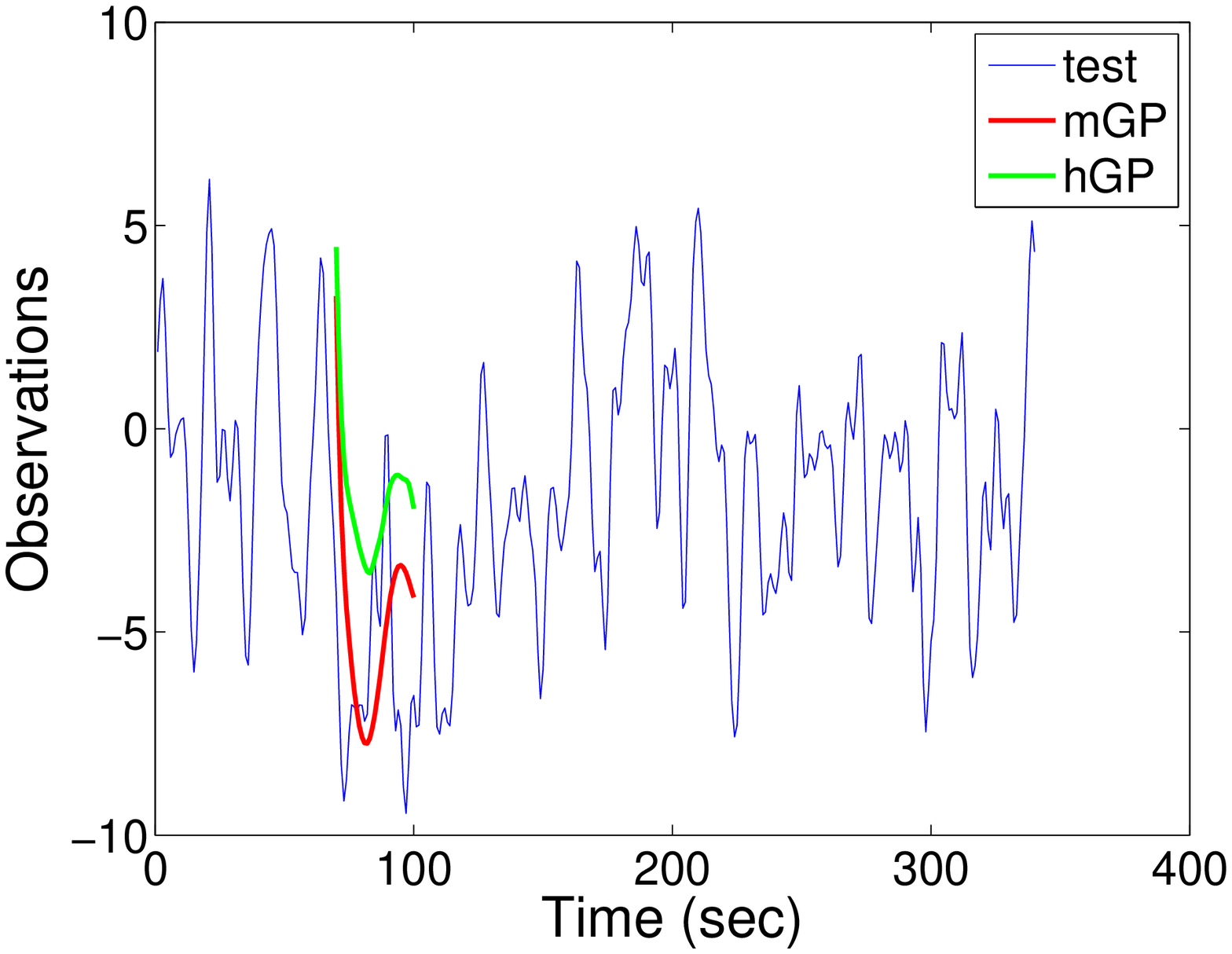}\\
				\includegraphics[height = 1.2in]{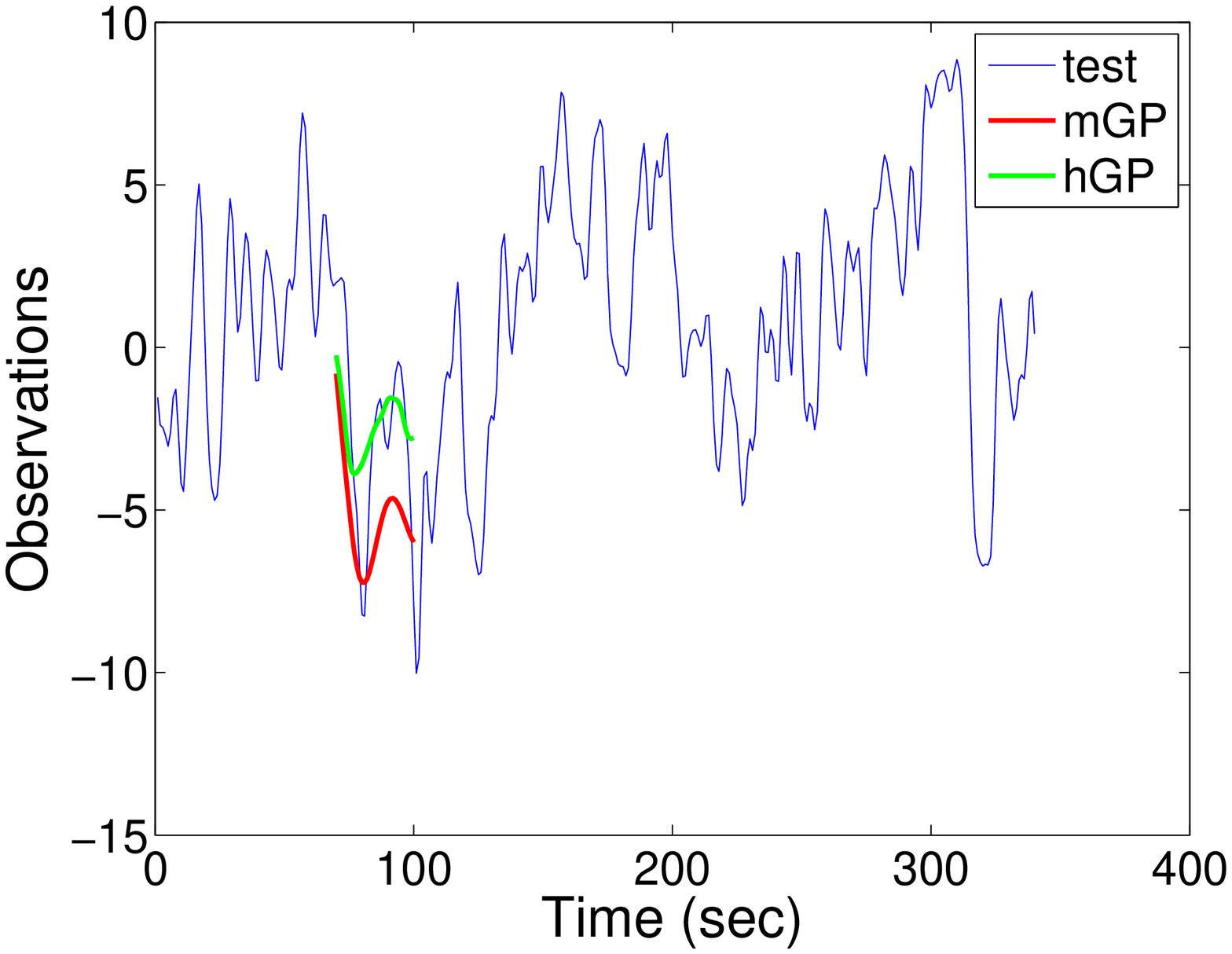} & \hspace{-0.01in}
				\includegraphics[height = 1.2in]{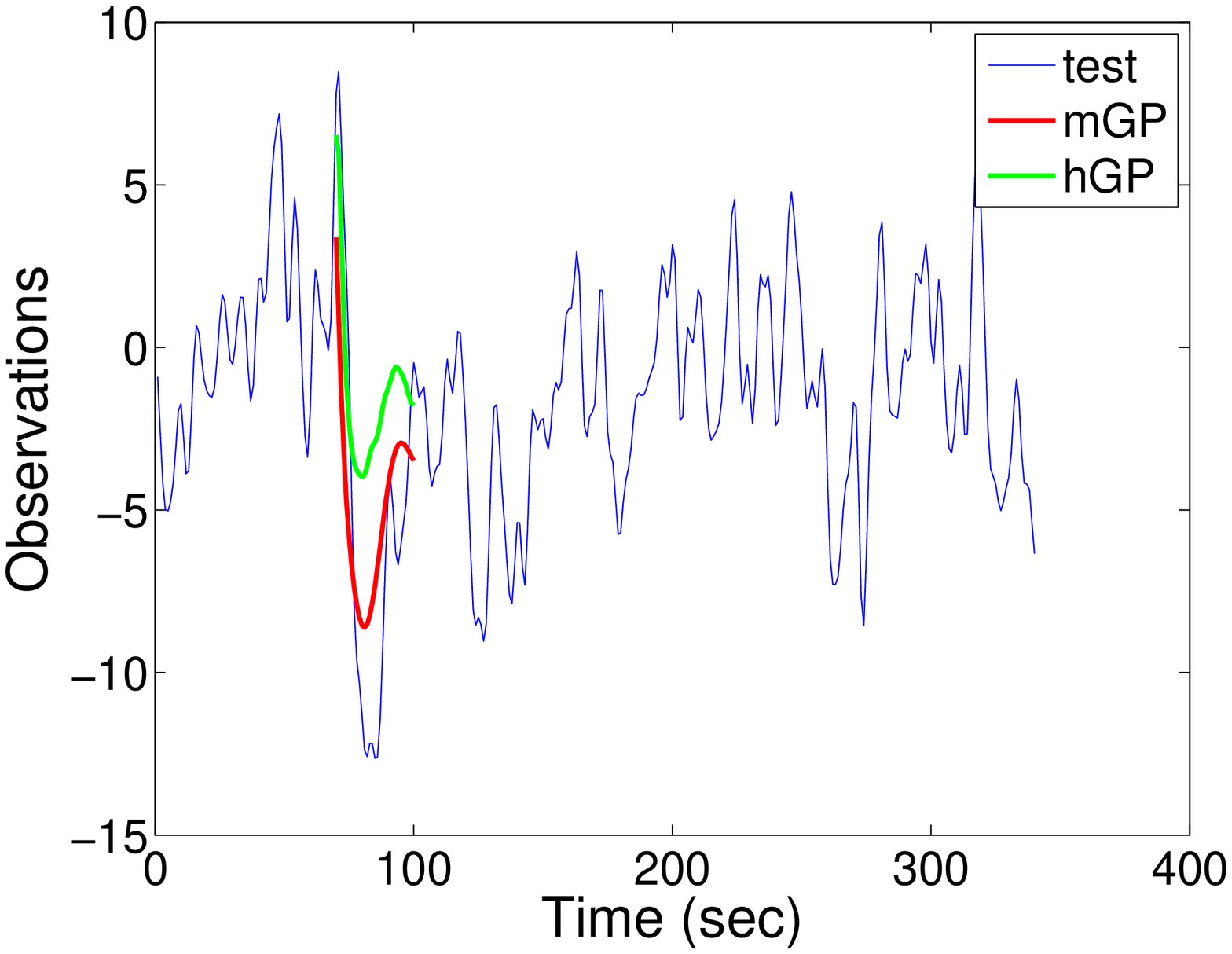} & \hspace{-0.1in}
				\includegraphics[height = 1.2in]{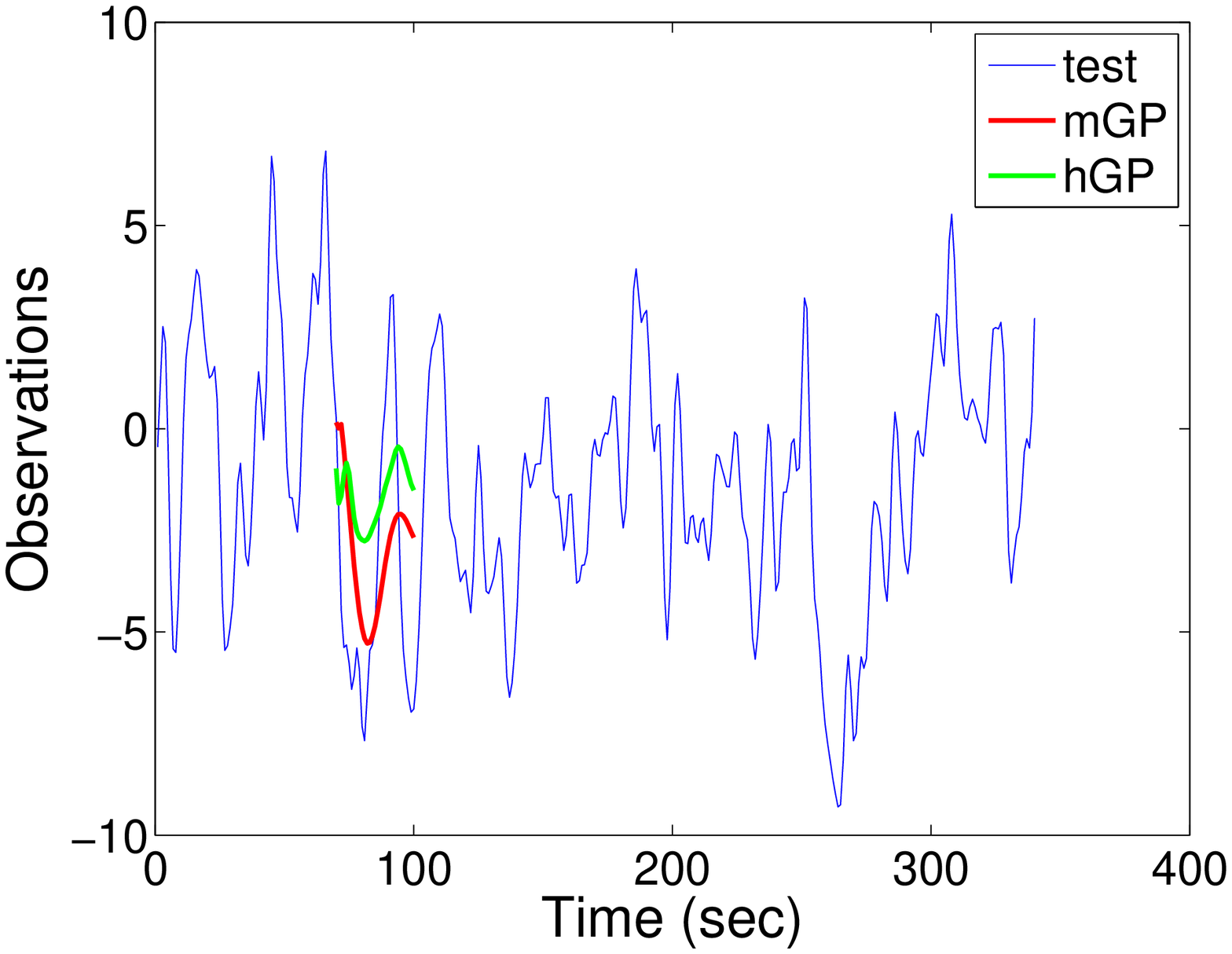}
		\end{tabular}
			\precap \vspace{-0.05in}
			\caption{\small Heldout test data for the visual cortex sensor 77 and 6 different words.  For $\tau=70$, we show the predictive mean $y^*_{\tau:\tau+30}$ under an hGP and mGP conditioned on $y^*_{1:\tau-1}$ and 15 training sequences.}
\label{fig:heldouttau}
\postcap
	\end{center} 
\end{figure}

In Fig.~\ref{fig:heldouttau}, we display examples of the heldout test data for the visual cortex sensor 77 and 6 different words.  Each plot only shows the first heldout trial even though the results of Fig. 7 perform a full analysis on each of the 5 heldout trials.  For $\tau=70$, we show the predictive mean $y^*_{\tau:\tau+30}$ conditioned on $y^*_{1:\tau-1}$ and 15 training sequences.  We compare the performance of the mGP to that of the hGP.
Only the predictive mean is displayed for clarity.  The predictive variances associated with the hGP were similar to, but slightly larger than those of the mGP (7\% larger on average).  The 95\% predictive intervals included the heldout observations in all 6 cases for the mGP and in 5 cases for the hGP.  However, note the significantly better mean predictions for the mGP.

\subsection{MEG Data Acquisition}

Subjects gave their written informed consent approved by University of Pittsburgh (protocol PRO09030355) and Carnegie Mellon (protocol HS09-343) Institutional Review Boards. MEG data were recorded using an Elekta Neuromag device (Elekta Oy).  While the machine has 306 sensors, to reduce the dimension of the data, only recordings from the second gradiometers were used for these experiments.  The data was acquired at 1 kHz, high-pass filtered at 0.1 Hz and low-pass filtered at 330 Hz. Eye movements (horizontal and vertical eye movements as well as blinks) were  monitored by recording the differential activity of muscles above below and beside the eyes.  At at the beginning of each session we recorded the position of the participants head with four head position indicator (HPI) coils placed on the subject's scalp. The HPI coils, along with three cardinal points (nasian, left and right pre-auricular), were digitized into the system. 

The data were preprocessed using the Signal Space Separation method (SSS)~\cite{Taulu2004,Taulu2006} and temporal extension of SSS (tSSS)~\cite{Taulu2009} to remove artifacts and noise unrelated to brain activity.  In addition, we used tSSS to realign the head position measured at the beginning of each block to a common location. The MEG signal was then low-pass filtered to 50 Hz to remove the contributions of line noise and down-sampled to 200 Hz. The Signal Space Projection method (SSP)~\cite{Uusitalo1997} was then used to remove signal contamination by eye blinks or movements, as well as MEG sensor 
malfunctions or other artifacts.  Each MEG repetition starts 260 ms before stimulus onset, and ends 1440 ms after stimulus onset, for a total of 1.7 seconds and 340 time points of data per sample.  MEG recordings are known to drift with time, so we correct our data by subtracting the mean signal amplitude during the 200ms before stimulus onset, for each sensor/repetition pair. Because the magnitude of the MEG signal is very small, we multiply the signal by $10^{12}$ to avoid numerical precision problems.

\section{MCMC Sampler Pseudocode}

We assume (i) a cost matrix $W$ formed from the absolute value of the empirical correlation matrix of a set of training replicates, (ii) a prior $F$ on the partition points, and (iii) hyperparameters $\theta = \{\kappa,d^0,\dots,d^{L-1},\sigma^2\}$ defining the mGP kernel bandwith, variances, and nugget noise, respectively.  The sampler is initialized with a hierarchical partition $\A$ drawn from the normalized cut proposal $q$.  The covariance matrix $\Sigma$ is a deterministic function of the hierarchical partition $\A$ and the hyperparameters $\theta$.  In what follows, we use $\texttt{kernel}$ to define a function that provides this mapping for a given partition set at a given tree level.  The likelihood $p(Y \mid \Sigma,\theta) = p(Y \mid \A,\theta)$ is computed exactly as in Eq. (13) of the main paper.  Algorithm~\ref{alg:samplerGlobal} details the global search iterations and Algorithm~\ref{alg:samplerLocal} the local (which can also produce global searches if the root note is selected).  The local search algorithm additionally assumes a node-proposal distribution indicated by $\texttt{nodeproposal}$.

\newpage

\begin{algorithm}[t]
   \caption{One Iteration of mGP MCMC Sampler - GLOBAL SEARCH}
   \label{alg:samplerGlobal}
\begin{algorithmic}
   \STATE {\bfseries Input:} Cost matrix $W$, input locations $\X$, hyperparameters $\theta$,\\
	\STATE \hspace{0.5in} previous partition $\A$ and corresponding $\Sigma$\algspace
   \STATE $\{z_1,\dots z_{2^{L-1}-1}\} \leftarrow $ partition points of $\A$\algspace
	\STATE $\A^{'0} = \X$, $\Sigma' = 0_{n \times n}$\hfill initialize structures for proposal\algspace
   \FOR{$\ell=1,\ldots,L-1$}\algspace
	\FOR{$\nu=1:2:2^\ell$}\algspace
   \STATE  $\{\A_\nu^{'\ell},\A_{\nu+1}^{'\ell}\} \sim q(\cdot \mid \A_{(\nu+1)/2}^{'\ell-1},W)$ \hfill normalized cut proposal \algspace
	\STATE $\Sigma'(\A_\nu^{'\ell}) = \Sigma'(\A_\nu^{'\ell}) + \texttt{kernel}(\A_\nu^{'\ell},\theta,\ell)$ \hfill add $K_\ell$ submatrix corresponding to $\A_{\nu}^{'\ell}$\algspace
	\STATE $\Sigma'(\A_{\nu+1}^{'\ell}) = \Sigma'(\A_{\nu+1}^{'\ell}) + \texttt{kernel}(\A_{\nu+1}^{'\ell},\theta,\ell)$\hfill add $K_\ell$ submatrix corresponding to $\A_{\nu+1}^{'\ell}$\algspace
   \ENDFOR
	\ENDFOR
	\STATE $\Sigma' = \Sigma' + \sigma^2I_n$\algspace
	\STATE $\{z'_1,\dots z'_{2^{L-1}-1}\} \leftarrow $ partition points of $\A'$\algspace
   \STATE $\rho \sim \mbox{Ber}(\min(r(\A' \mid \A),1))$, \quad $r(\A' \mid \A) = \frac{p(Y \mid \Sigma',\theta)\prod_i F(z'_i) \prod_{\nu_{odd},\ell} q(\{\A_\nu^{\ell},\A_{\nu+1}^{\ell}\} \mid \A_{(\nu+1)/2}^{\ell-1},W)}
{p(Y \mid \Sigma,\theta)\prod_i F(z_i) \prod_{\nu_{odd},\ell} q(\{\A_\nu^{'\ell},\A_{\nu+1}^{'\ell}\} \mid \A_{(\nu+1)/2}^{'\ell-1},W)}$\algspace
	\STATE $\A \leftarrow \rho\A' + (1-\rho)\A$, \quad $\Sigma \leftarrow \rho\Sigma' + (1-\rho)\Sigma$\hfill accept or reject proposal\algspace
   \STATE  {\bfseries Output:} $\A,\Sigma$
\end{algorithmic}
\end{algorithm}

\begin{algorithm}[h!]
   \caption{One Iteration of mGP MCMC Sampler - LOCAL SEARCH}
   \label{alg:samplerLocal}
\begin{algorithmic}
   \STATE {\bfseries Input:} Cost matrix $W$, input locations $\X$, hyperparameters $\theta$,\\
	\STATE \hspace{0.5in} previous partition $\A$ and corresponding $\Sigma$\algspace
   \STATE $\{z_1,\dots z_{2^{L-1}-1}\} \leftarrow $ partition points of $\A$\algspace
	\STATE $\Sigma' \leftarrow \Sigma$, \quad $\A' \leftarrow \A$ \hfill initialize proposals to previous values\algspace
	\STATE $\A_{\nu*}^{\ell*} \sim \texttt{nodeproposal}$ \hfill select a set (tree node) to repartition\algspace
	\STATE $S \leftarrow \{(\nu,\ell)\mid \A_{\nu}^{'\ell} \subset \A_{\nu*}^{\ell*}\}$ \hfill node descendants\algspace
	\FOR{$(\nu,\ell) \in S$}
	\STATE $\Sigma'(\A_{\nu}^{'\ell}) = \Sigma'(\A_{\nu}^{'\ell}) - \texttt{kernel}(\A_{\nu}^{'\ell},\theta,\ell)$ \hfill remove contributions from node descendants\algspace
	\ENDFOR
	\FOR{$(\nu,\ell) \in S$ such that $\nu$ is odd}
   \STATE  $\{\A_\nu^{'\ell},\A_{\nu+1}^{'\ell}\} \sim q(\cdot \mid \A_{(\nu+1)/2}^{'\ell-1},W)$ \hfill normalized cut proposal \algspace
	\STATE $\Sigma'(\A_\nu^{'\ell}) = \Sigma'(\A_\nu^{'\ell}) + \texttt{kernel}(\A_\nu^{'\ell},\theta,\ell)$ \hfill add $K_\ell$ submatrix corresponding to $\A_{\nu}^{'\ell}$\algspace
	\STATE $\Sigma'(\A_{\nu+1}^{'\ell}) = \Sigma'(\A_{\nu+1}^{'\ell}) + \texttt{kernel}(\A_{\nu+1}^{'\ell},\theta,\ell)$\hfill add $K_\ell$ submatrix corresponding to $\A_{\nu+1}^{'\ell}$\algspace
   \ENDFOR
	\STATE $\{z'_1,\dots z'_{2^{L-1}-1}\} \leftarrow $ partition points of $\A'$\algspace
   \STATE $\rho \sim \mbox{Ber}(\min(r(\A' \mid \A),1))$, \hfill $r(\A' \mid \A) = \frac{p(Y \mid \Sigma',\theta)\prod_i F(z'_i) \prod_{(\nu_{odd},\ell) \in S} q(\{\A_\nu^{\ell},\A_{\nu+1}^{\ell}\} \mid \A_{(\nu+1)/2}^{\ell-1},W)}
{p(Y \mid \Sigma,\theta)\prod_i F(z_i) \prod_{(\nu_{odd},\ell) \in S} q(\{\A_\nu^{'\ell},\A_{\nu+1}^{'\ell}\} \mid \A_{(\nu+1)/2}^{'\ell-1},W)}$\algspace
	\STATE $\A \leftarrow \rho\A' + (1-\rho)\A$, \quad $\Sigma \leftarrow \rho\Sigma' + (1-\rho)\Sigma$\hfill accept or reject proposal\algspace
   \STATE  {\bfseries Output:} $\A,\Sigma$
\end{algorithmic}
\end{algorithm}

\newpage
\bibliographystyle{plainnat} \presec \small 
\bibliography{../../Bibliography/Bibliography,../../Bibliography/Bibliography_BPARHMM}

\end{document}